\documentclass{appolb}
\usepackage{epsfig}

\preprint{~}

\newcommand{\Dphi}{$\Delta \phi$}
\newcommand{\Deta}{$\Delta \eta$}
\newcommand{\DR}{$\Delta R$}

\newcommand{\pt}{$p_T$}
\newcommand{\kt}{$k_T$}
\newcommand{\ttbar}{t$\bar{\rm t}$}
\newcommand{\bbbar}{b$\bar{\rm b}$}
\newcommand{\ccbar}{c$\bar{\rm c}$}
\newcommand{\invpb}{pb$^{-1}$}
\newcommand{\invfb}{fb$^{-1}$}
\newcommand{\gevc}{GeV/$c$}
\newcommand{\Zg}{Z/$\gamma^*$}

\def\slashchar#1{\setbox0=\hbox{$#1$}           
   \dimen0=\wd0                                 
   \setbox1=\hbox{/} \dimen1=\wd1               
   \ifdim\dimen0>\dimen1                        
      \rlap{\hbox to \dimen0{\hfil/\hfil}}      
      #1                                        
   \else                                        
      \rlap{\hbox to \dimen1{\hfil$#1$\hfil}}   
      /                                         
   \fi}
\def\MET{$\slashchar{E}_T$}

\begin{document}
\title{QCD and Top-Quark Results from the Tevatron
\thanks{Presented at the Physics at LHC 2006 Conference for the CDF and D0 Collaborations}%
}
\author{Marek Zieli\'nski
\address{University of Rochester, Rochester, NY}
}
\maketitle
\begin{abstract}
Selected recent QCD and top-quark results from the Tevatron
are reviewed, aiming to illustrate progression from basic studies
of QCD processes to verification of perturbative calculations and 
Monte Carlo simulation tools, and to their applications 
in more novel and complex cases, like top-quark studies and
searches for new physics.
 
\end{abstract}
\PACS{12.38.Qk, 13.85.Rm, 13.87.-a, 14.65.Ha}
  
\section{QCD Results}

QCD processes provide signals to test theoretical calculations
and models and contribute major backgrounds to many
measurements. Thus, their detailed understanding and modeling
is of crucial importance.

\begin{figure}[!Hhtb]
\centerline{
\epsfxsize=2.5in
\epsffile{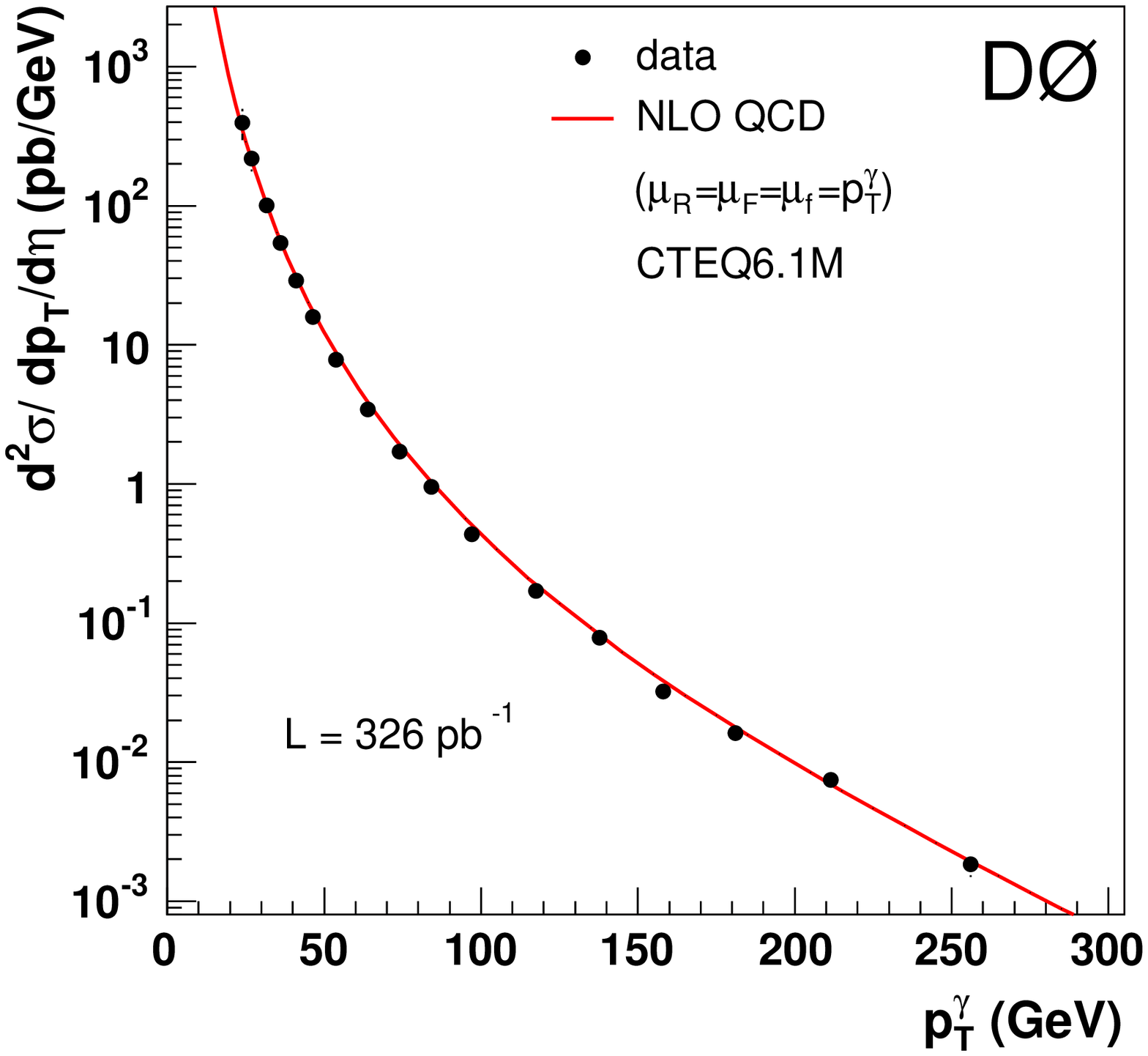}
\epsfxsize=2.5in
\epsffile{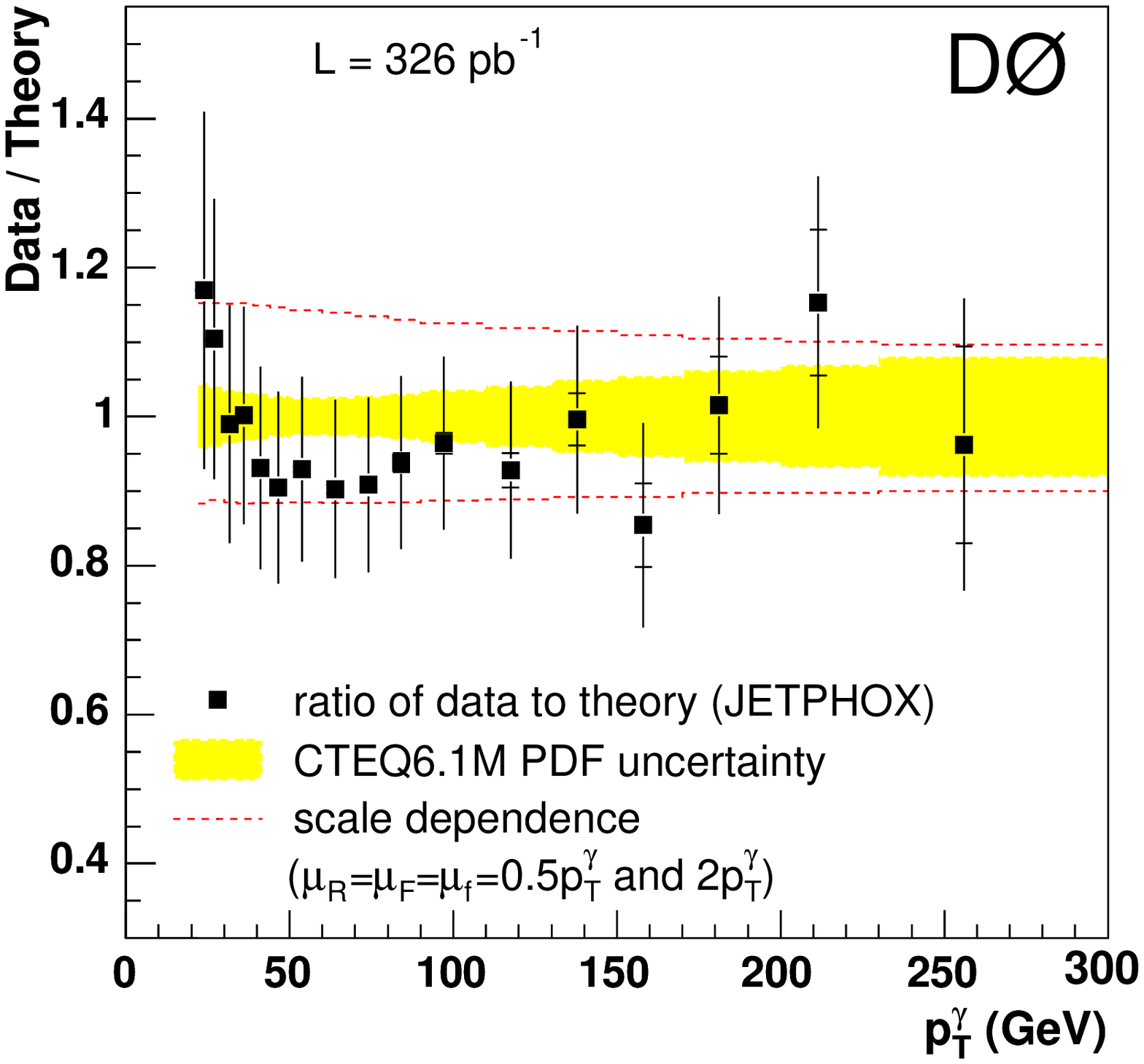}
}
\caption{
Left: Inclusive isolated-$\gamma$ cross section vs \pt.
Right: Ratio to NLO pQCD. 
\label{fig:d0-photons}}
\end{figure}

Production of isolated photons at large \pt\ provides
one of the cleanest and most accurate tests of perturbative QCD (pQCD).
Such photons originate primarily from hard collisions of partons
(quark-gluon or quark-antiquark) and are thus sensitive to the
parton distribution functions (PDFs). Consequently, they can help
constrain PDFs (especially the large-$x$ gluon distribution)
independently of the high-\pt\ jet production. Such constraints can
reduce ambiguities in interpreting results on high-\pt\
jet production in terms of new physics.
Isolated photon samples also provide indispensable calibration of
the recoiling jets.
D0 presented the first measurement of the inclusive isolated
photon cross section in Run 2 of the Tevatron \cite{d0-photons}.
The photon spectrum, obtained using 326 \invpb, 
spans \pt = 23--300 \gevc\ and $|\eta|<0.9$,
significantly extending the reach observed in Run 1. 
The measurement agrees well with next-to-leading order (NLO)
pQCD calculation \cite{jetphox}
over six orders of magnitude, Fig.~\ref{fig:d0-photons}(left).
The data/theory ratio, presented in the right panel,
shows that the theoretical scale dependence
and PDF uncertainties are comparable to the experimental
error bars. Further improvements in theoretical predictions
are desired to reduce the level of sensitivity to the choice of pQCD
scales in order to fully exploit the potential of this measurement
for constraining PDFs with the help of much larger data samples 
already available.

\begin{figure}[!Hhtb]
\centerline{
\epsfxsize=2.in
\epsffile{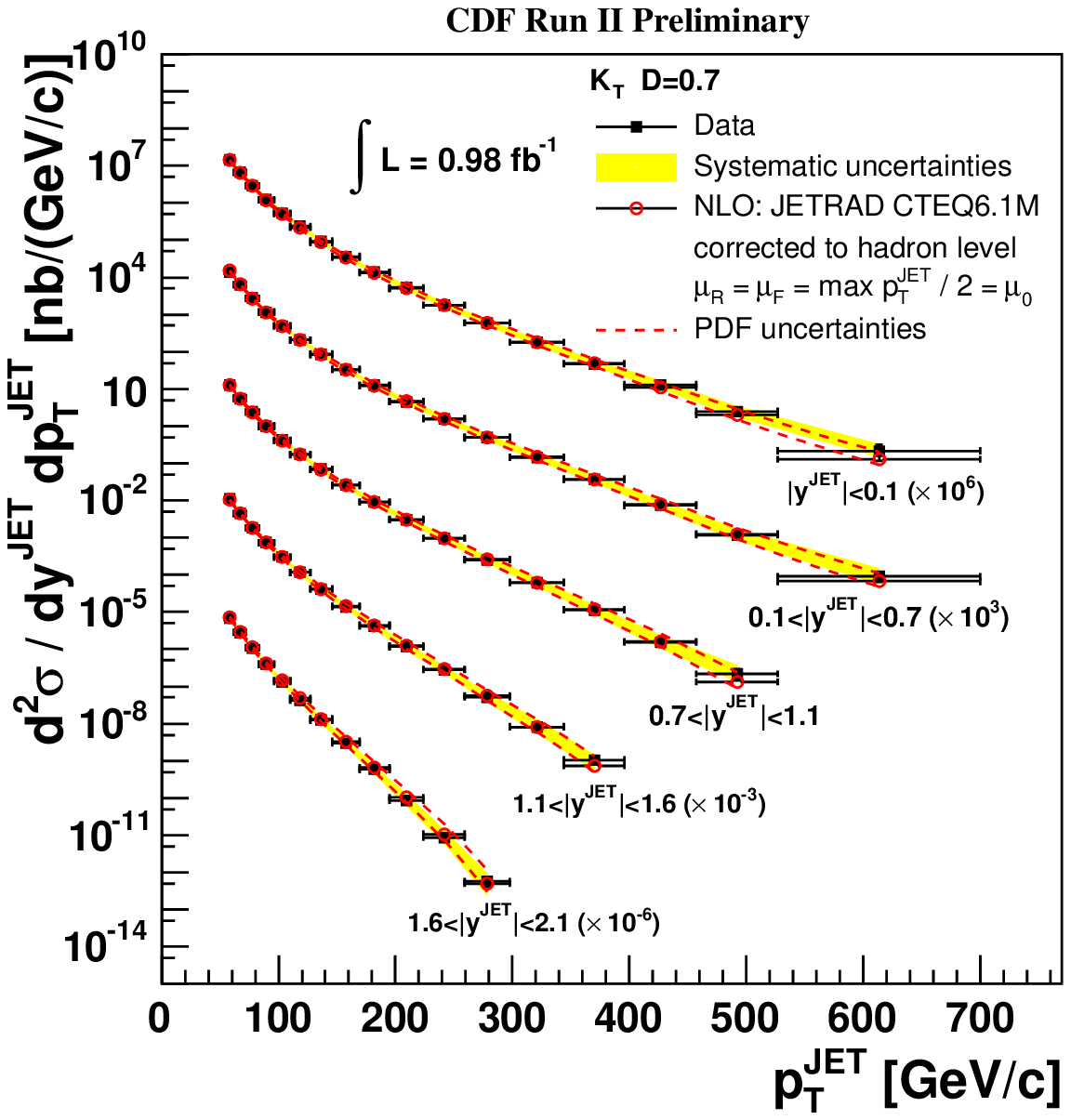}
\epsfxsize=3in
\epsffile{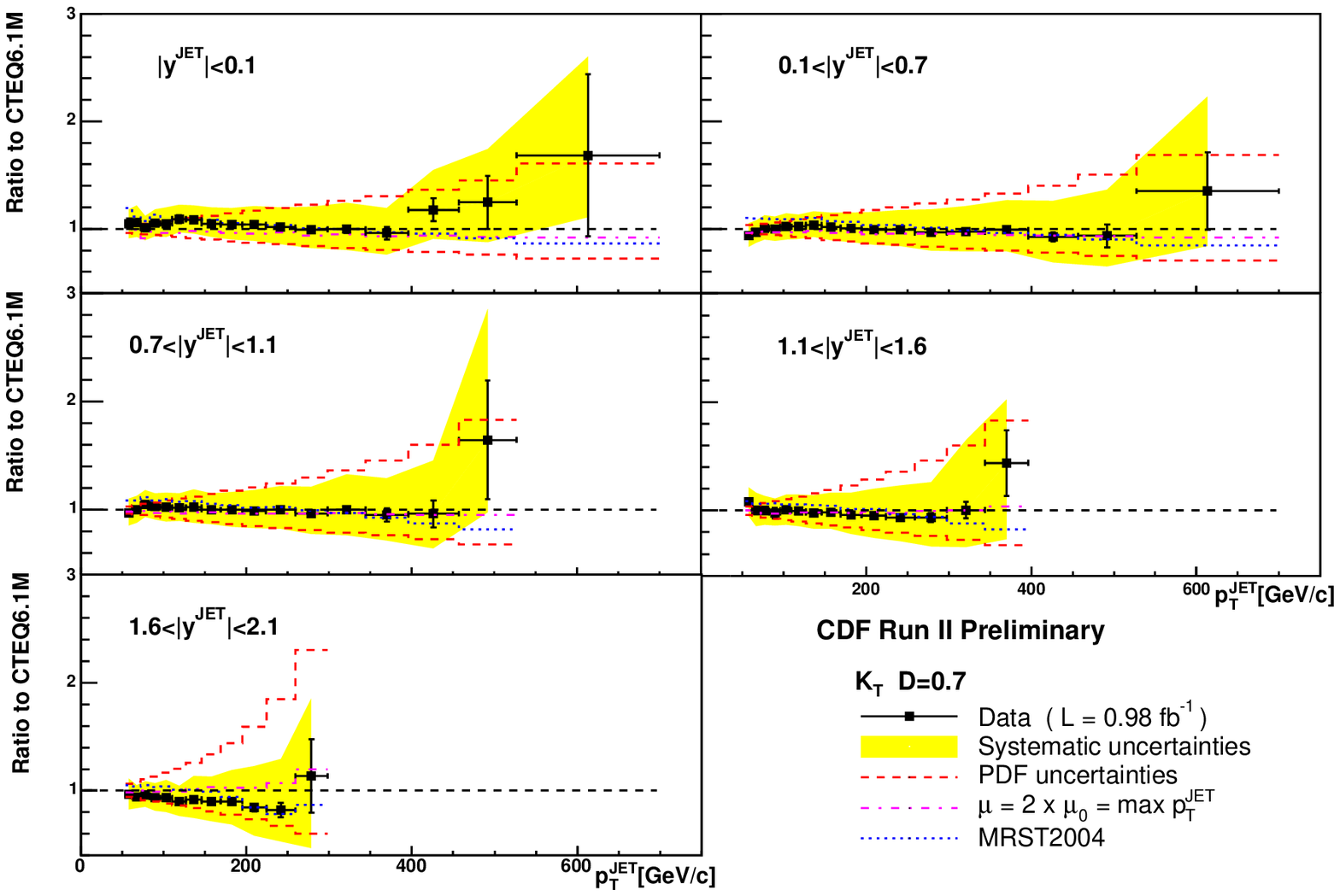}
}
\caption{
Left: Inclusive \kt-jet cross section vs \pt.
Right: Ratio to NLO pQCD.
\label{fig:cdf-kt}}
\end{figure}

Measurements of the inclusive jet cross section provide tests of pQCD
and sensitivity to new physics by probing distances down to $\approx 10^{-19}$ m.
Results at large rapidities are particularly important
for constraining PDFs in a kinematic region where no effects from
new physics are expected.
CDF obtained the first measurement in Run 2 of inclusive jet cross section
in five rapidity regions using the longitudinally-invariant \kt\ algorithm
and $\approx 1$ \invfb\ of data \cite{cdf-kt}.
Figure~\ref{fig:cdf-kt}(left) shows the results for the size parameter
$D=0.7$ for jets with \pt$>54$ \gevc\ and $|y|<2.1$. The right panel shows
the data ratio to NLO theory 
\cite{jetrad}
and displays the experimental and theoretical 
uncertainties. The former are dominated by the jet energy calibration
and the latter by the QCD scale and PDF variations. The theoretical
calculations include corrections for 
non-perturbative effects related to the underlying event and 
hadronization process. These corrections are essential to obtain 
good agreement between data and theory.
Similar level of  agreement has been found  for central jets
($0.1<|y|<0.7$) using values of $D=0.5$ and 1.0 and the corresponding
corrections. These results demonstrate veracity of the \kt\ algorithm
in the hadron-collider environment within the range of the
measurement. For the most forward
rapidity bin ($1.6<|y|<2.1$) the experimental uncertainty is smaller than
the one due to PDFs, hence this measurement is expected to further constrain
large-$x$ PDFs in future global fits. 
Similar conclusions have been reached by CDF and D0 for
the inclusive jet cross section measurements using the MidPoint cone
algorithm (not shown). Here corrections
for soft effects are smaller than for \kt\ algorithm but non-negligible at
the current level of precision.

\begin{figure}[!Hhtb]
\centerline{
\epsfxsize=2.in
\epsffile{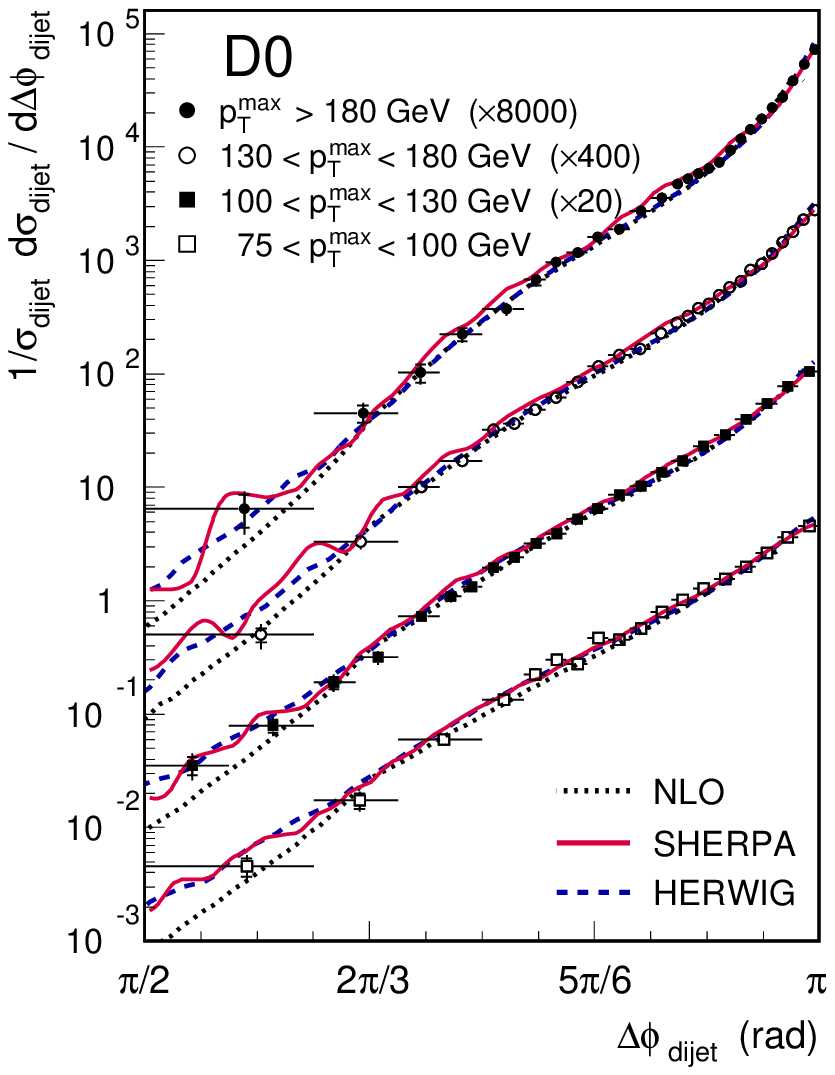}
\epsfxsize=2.in
\epsffile{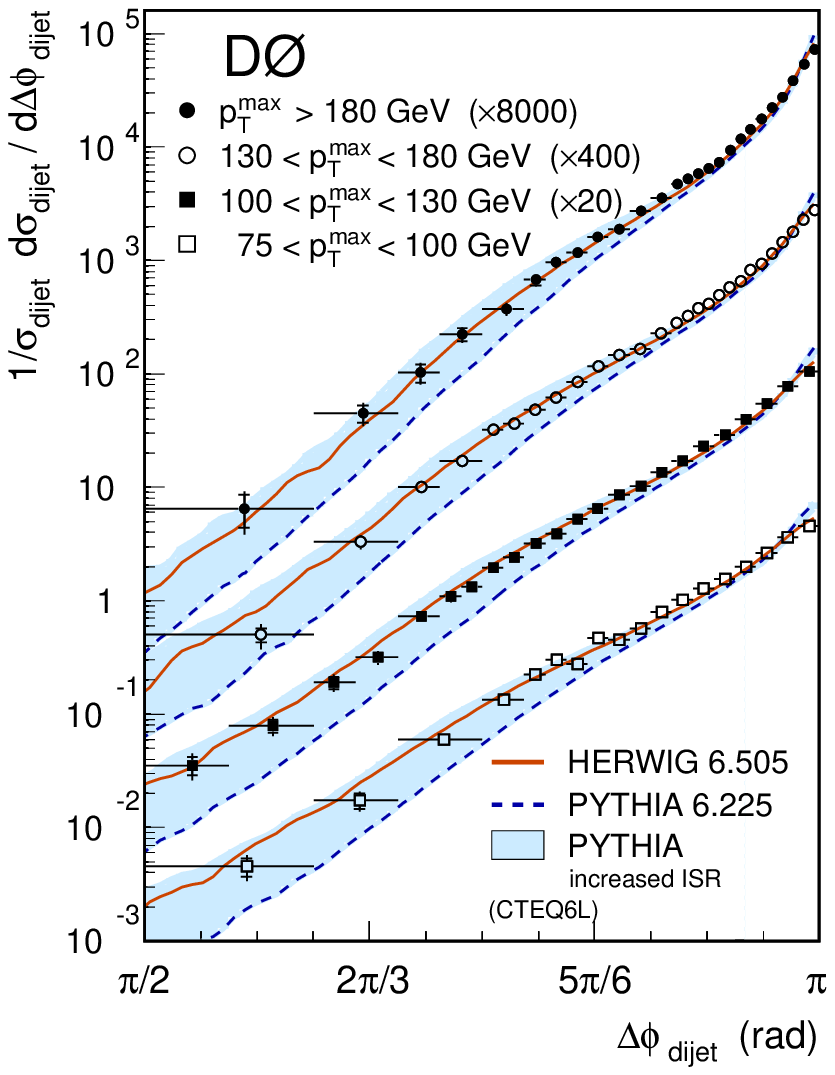}
}
\caption{
Left: dijet \Dphi\ distributions compared to NLO pQCD, {\sc herwig}
and {\sc sherpa}. 
Right: Comparison to {\sc pythia} with varied Initial State Radiation (ISR).
\label{fig:d0-deltaphi}}
\end{figure}

Correlations in the azimuthal angle \Dphi\ between the two 
leading jets
in an event provide a clean and simple probe of radiation effects.
In the absence of radiation \Dphi = $\pi$. Soft radiation causes
small deviations from $\pi$ while \Dphi\ significantly lower
than $\pi$ indicates the presence of hard radiation, such as
additional jets with high \pt. 
The proper description of multi-parton radiation
is crucial for a wide range of precision measurements as
well as for searches for new physical phenomena at 
Tevatron and LHC.
D0 results \cite{dphi} for \Dphi\ correlations between central
jets ($|\eta|<0.5$) are 
presented in Fig.~\ref{fig:d0-deltaphi} 
in four ranges of leading-jet \pt.
Since the data 
are sensitive to a range of jet multiplicities, they provide a test of
recent Monte Carlo approaches that combine
exact LO pQCD matrix elements for multi-parton production with
parton-shower models and of the associated ``matching'' prescriptions
imposed to avoid double-counting of equivalent parton configurations.
Two such generators, {\sc alpgen}
~\cite{alpgen}
(not shown) and {\sc sherpa}
~\cite{sherpa},
are in good agreement with data,
thus enhancing confidence in their applications to other processes.
The data are also well described by  NLO pQCD for three-jet production
~\cite{nlojet},
and by {\sc herwig} 
\cite{herwig} 
with default parameters.
Distributions from {\sc pythia} 
\cite{pythia} 
are  sensitive to
the value of a parameter which controls the maximum allowed virtuality in the 
initial-state shower.
The shaded bands in Fig.~\ref{fig:d0-deltaphi}(right) show the range of 
predictions when this parameter
is varied by a factor of four. The optimal value of 2.5 has been
incorporated in the recent tunes DW and DWT of {\sc pythia} parameters \cite{tune-dw}. 

\begin{figure}[!Hhtb]
\centerline{
\epsfxsize=2.5in
\epsffile{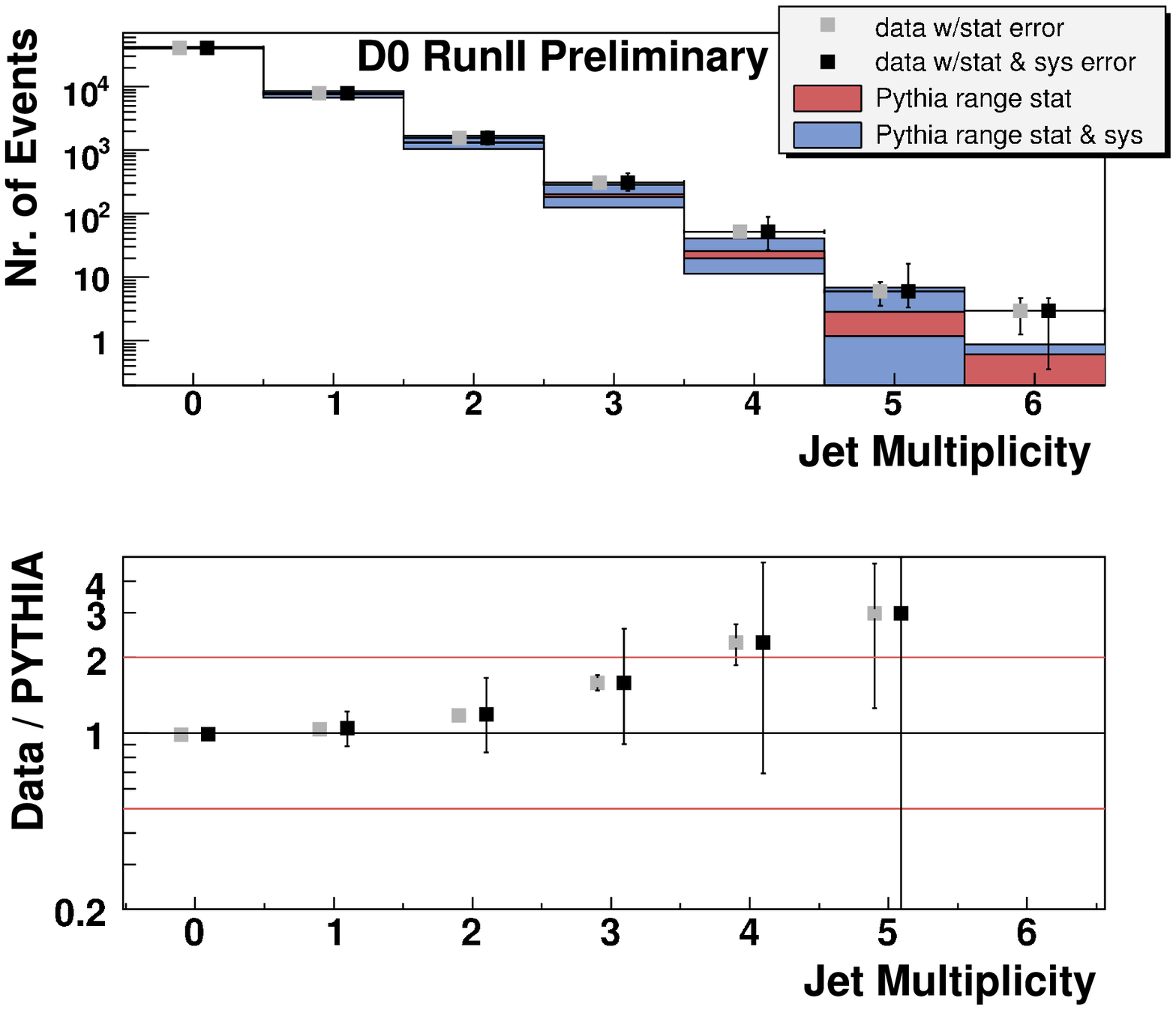}
\epsfxsize=2.5in
\epsffile{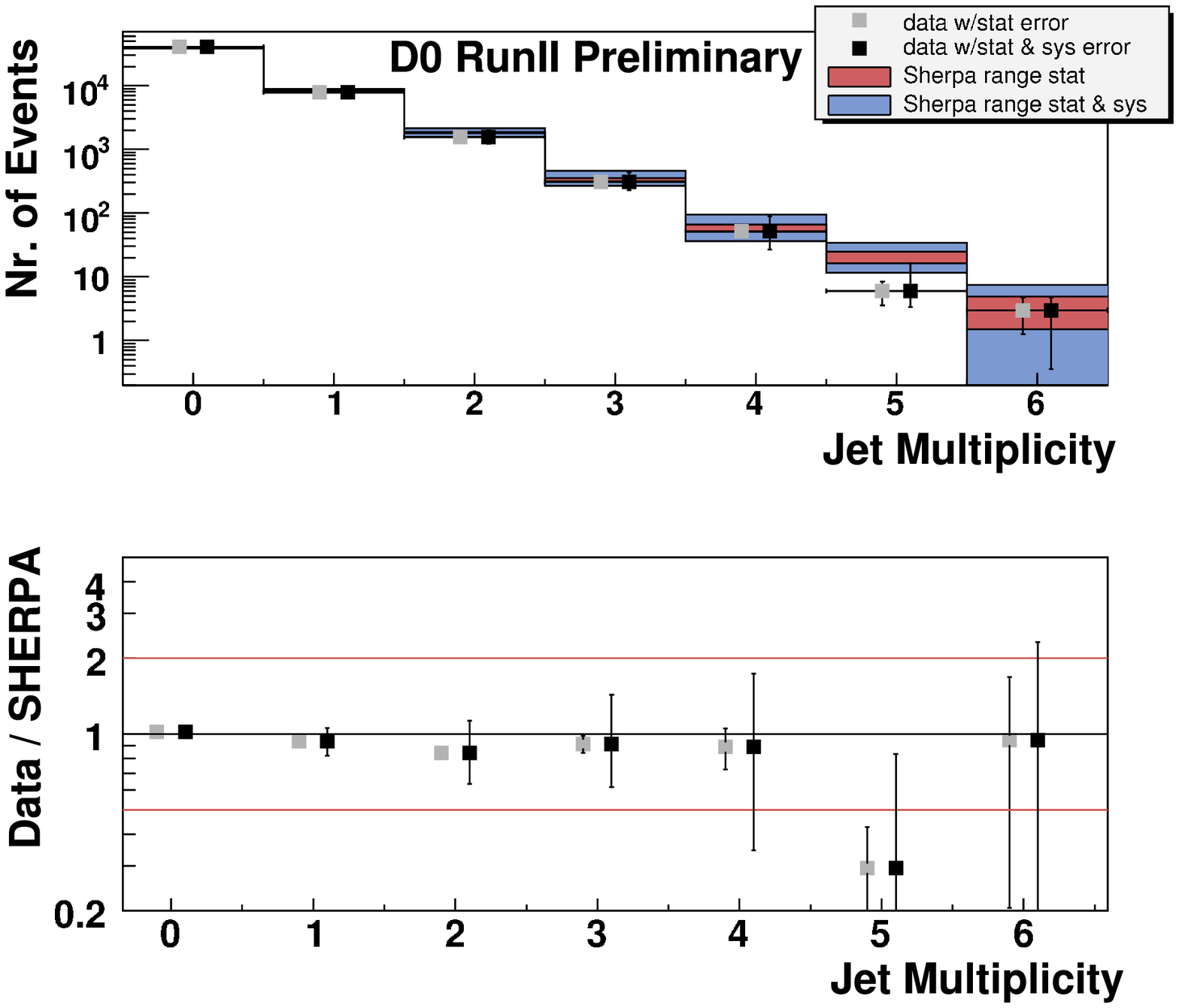}
}
\caption{Jet multiplicity in Z+jets events compared
to {\sc pythia}  and {\sc sherpa}.
\label{fig:d0-zjets}}
\end{figure}

Production of W and Z bosons in association with jets constitutes an important 
background to top-quark production and in the searches for new
physics, including production of the Higgs boson and
supersymmetric particles. Thus an accurate modeling of this
process is essential. The presence of W/Z ensures high $Q^2$
and facilitates tests of pQCD and Monte Carlo tools for
configurations with multiple soft jets.
D0 compared predictions from {\sc pythia} and {\sc sherpa}
to various distributions in \Zg+jets events using 950 \invpb
of data \cite{d0-zjets}.
Data selection required two electrons with \pt$>25$ \gevc\ and
$|\eta|<2.5$ within a di-electron mass window of 70--100 GeV, and
jets with \pt$>15$ \gevc.
{\sc pythia} was found to underestimate the 
production rate of higher jet multiplicities, Fig.~\ref{fig:d0-zjets}(left).
{\sc sherpa} provides a good description of jet multiplicity (right panel), 
and all kinematic distributions studied, including \pt\ distributions of the Z
and of 1$^{\rm st}$, 2$^{\rm nd}$ and 3$^{\rm rd}$ leading jets, 
as well as \Dphi\ and \Deta\ angular distributions between the jets.
Significant differences with data have been observed for {\sc pythia}
distributions.

\begin{figure}[!Hhtb]
\centerline{
\epsfxsize=2.5in
\epsffile{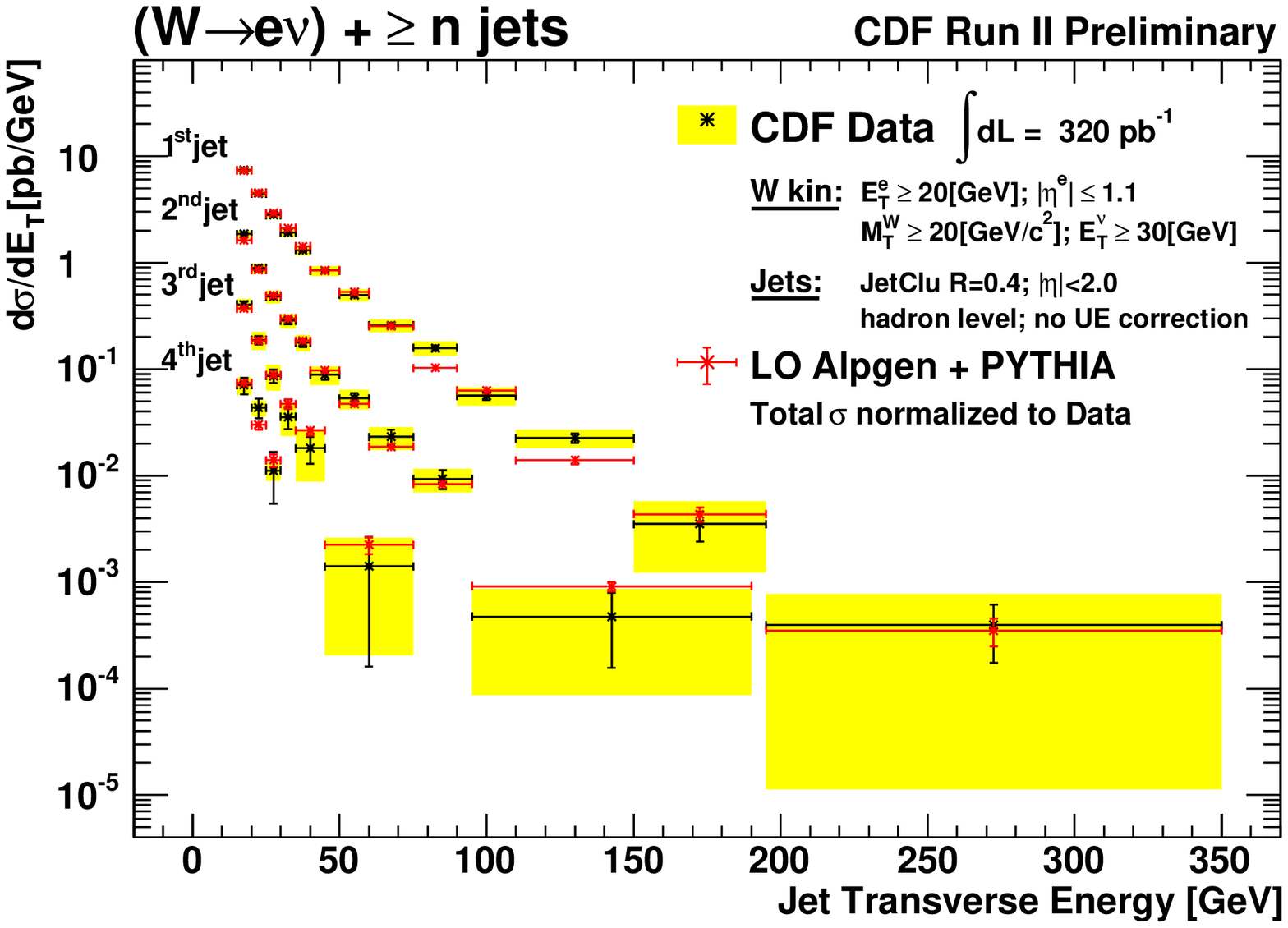}
\epsfxsize=2.5in
\epsffile{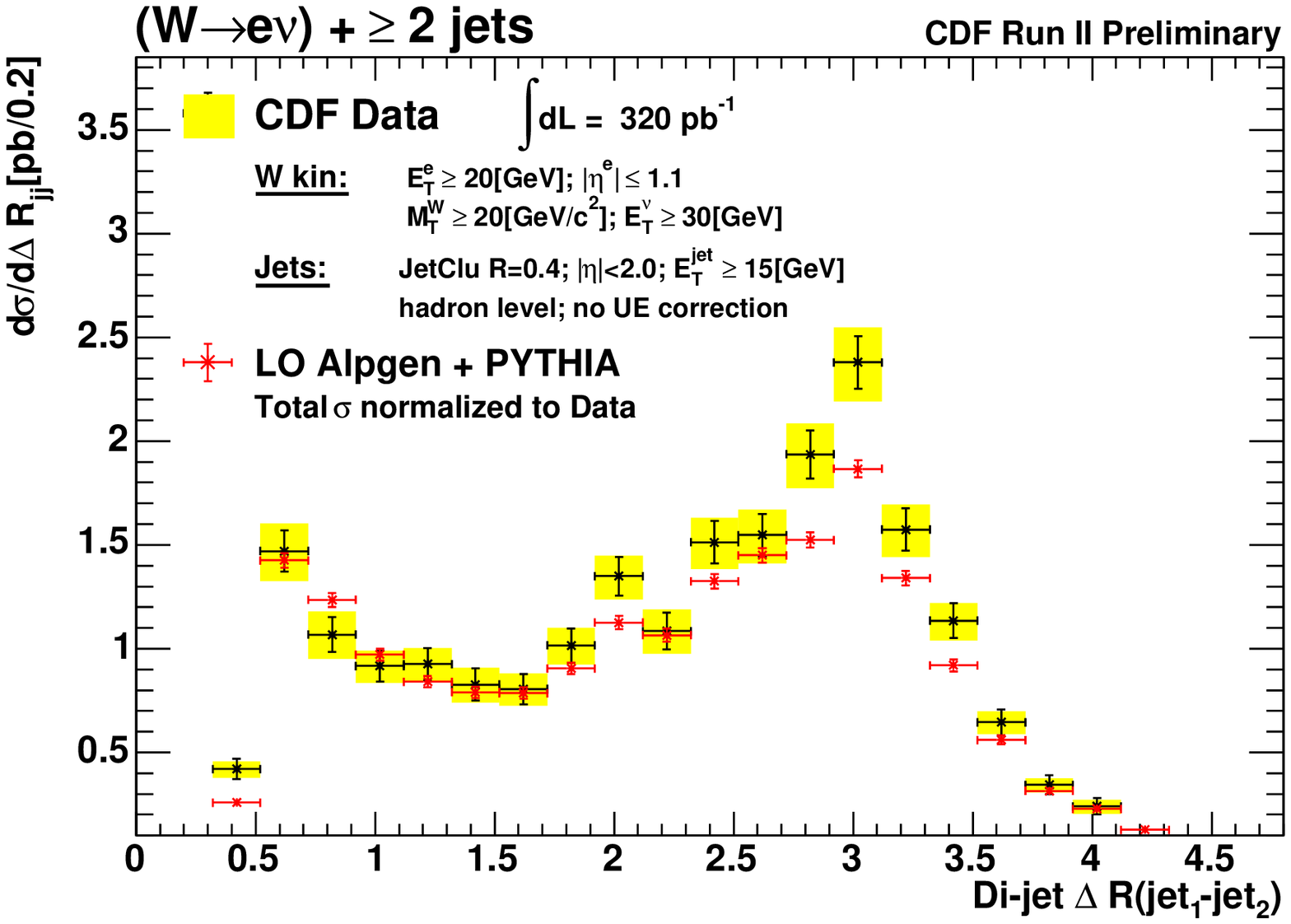}
}
\caption{Jet \pt\ and \DR\ in W+jets events compared to {\sc alpgen}.
\label{fig:cdf-wjets}}
\end{figure}

Using 320 \invpb\ of data
CDF performed \cite{cdf-wjets} shape comparisons between W+jets production
(up to four jets) with predictions from {\sc alpgen}
interfaced to {\sc pythia} for showering and hadronization.
Jets were corrected to hadron
level and kinematic cuts imposed to reduce model dependence
on acceptance and efficiency.  
Data selection required a good-quality electron candidate
with \pt$>20$ \gevc, missing transverse energy \MET $>30$ GeV, 
and $R = 0.4$ cone jets
with \pt$> 15$ \gevc\ and $|\eta|<2$. 
Reasonable agreement is observed for the jet \pt\ distributions
(Fig.~\ref{fig:cdf-wjets}(left)), \DR\ between jets in W+2jets
sample (Fig.~\ref{fig:cdf-wjets}(right)), 
and dijet invariant mass (not shown).

\section{Top-Quark Results}

Ten years after its discovery top quark is intensely studied
at the Tevatron. Its surprisingly large mass makes it the only
fermion having the Yukawa coupling near unity implying
its large contribution to the radiative corrections
to the Higgs mass. This leads to speculation that electroweak
symmetry breaking mechanism may be probed through studies
of its production and properties. Consequently, every aspect of
top-quark physics experimentally accessible is vigorously
scrutinized at the Tevatron.

The Standard Model (SM) predicts that at the Tevatron
top quarks are primarily produced in pairs
through the strong force by ${\rm q\bar{q}}$ annihilation 85\% of the time 
and by gg fusion 15\% of the time. The predicted cross section is
$\sigma_{\rm t\bar{t}} = 6.77 \pm 0.42$ pb for $m_{\rm t} = 175$ GeV
\cite{top-xs-theory}. In SM, top quarks decay $\approx$100\% to Wb, and hence
\ttbar\ events are classified according to the decay modes of the W's.
In dilepton events both W's decay into e or $\mu$.
This channel has a low branching fraction ($\approx 5$\%) but is very clean.
A recent extension of the dilepton analysis selects candidate events
requiring an isolated track instead of one of the leptons. 
This improves selection efficiency and enlarges the event sample
at the cost of additional backgrounds.
The channel  when one W decays to e or $\mu$ and the other to quarks
is called lepton+jets. It has a higher branching fraction ($\approx 30$\%)
but also receives higher backgrounds. Since it provides a large
but still fairly pure sample of top quarks, it facilitates
some of the best measurements in top physics.
Decays of both W's to quarks result in the all-hadronic channel,
which has the largest branching fraction ($\approx 44$\%) but also
the highest background from QCD multi-jet production.
The b-tagging information is essential for background suppression in this channel.
It also helps to reduce backgrounds in
the lepton+jets channel. Analyses based on decay modes involving $\tau$ 
leptons are especially difficult and are only now becoming developed.
Due to the presence of W's and jets in top decays, good understanding and 
simulation of QCD W/Z+jets and multijet production
is indispensable in top-quark measurements.

\begin{figure}[!Hhtb]
\centerline{
\epsfxsize=2.3in
\epsffile{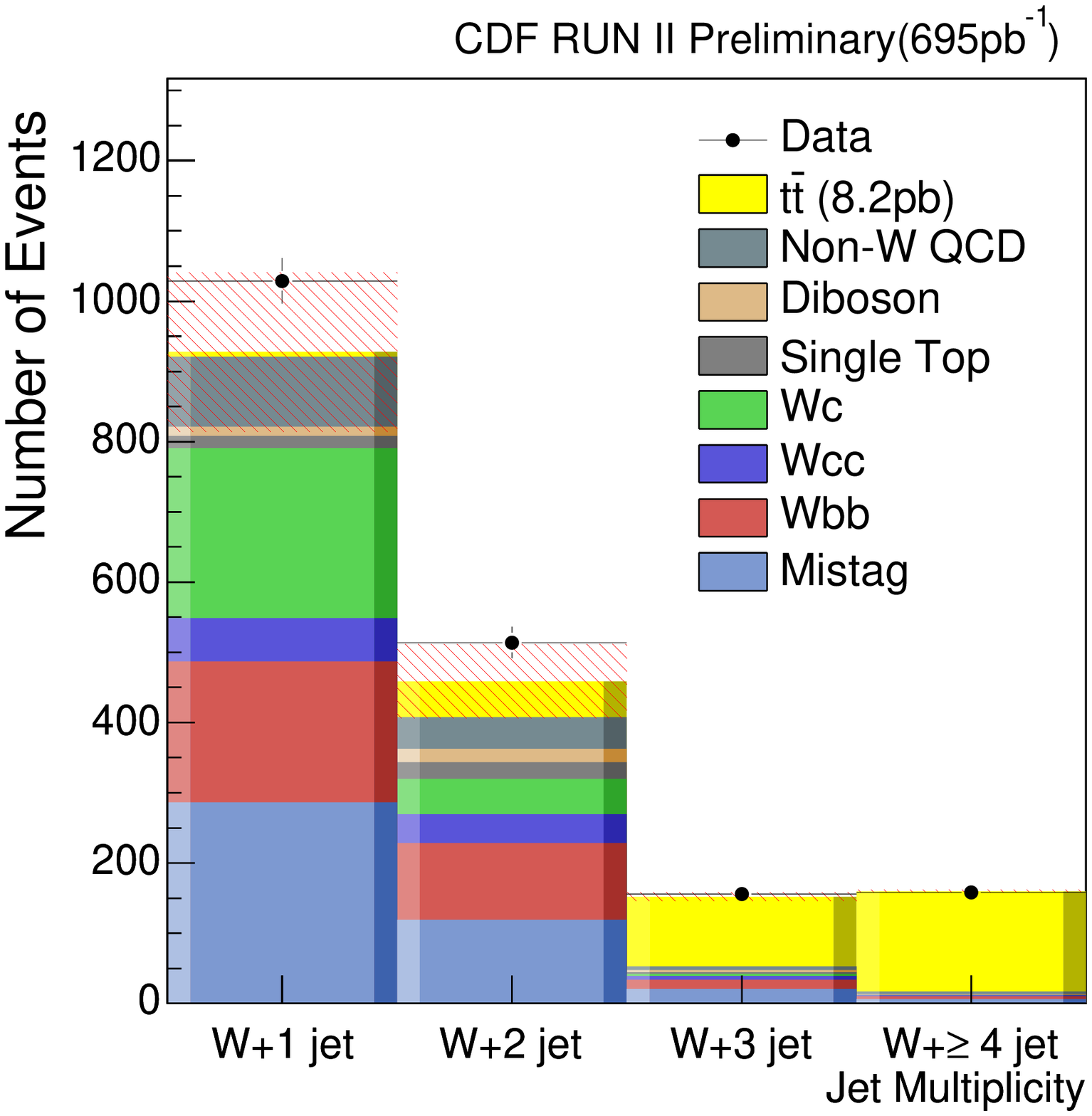}
\epsfxsize=2.3in
\epsffile{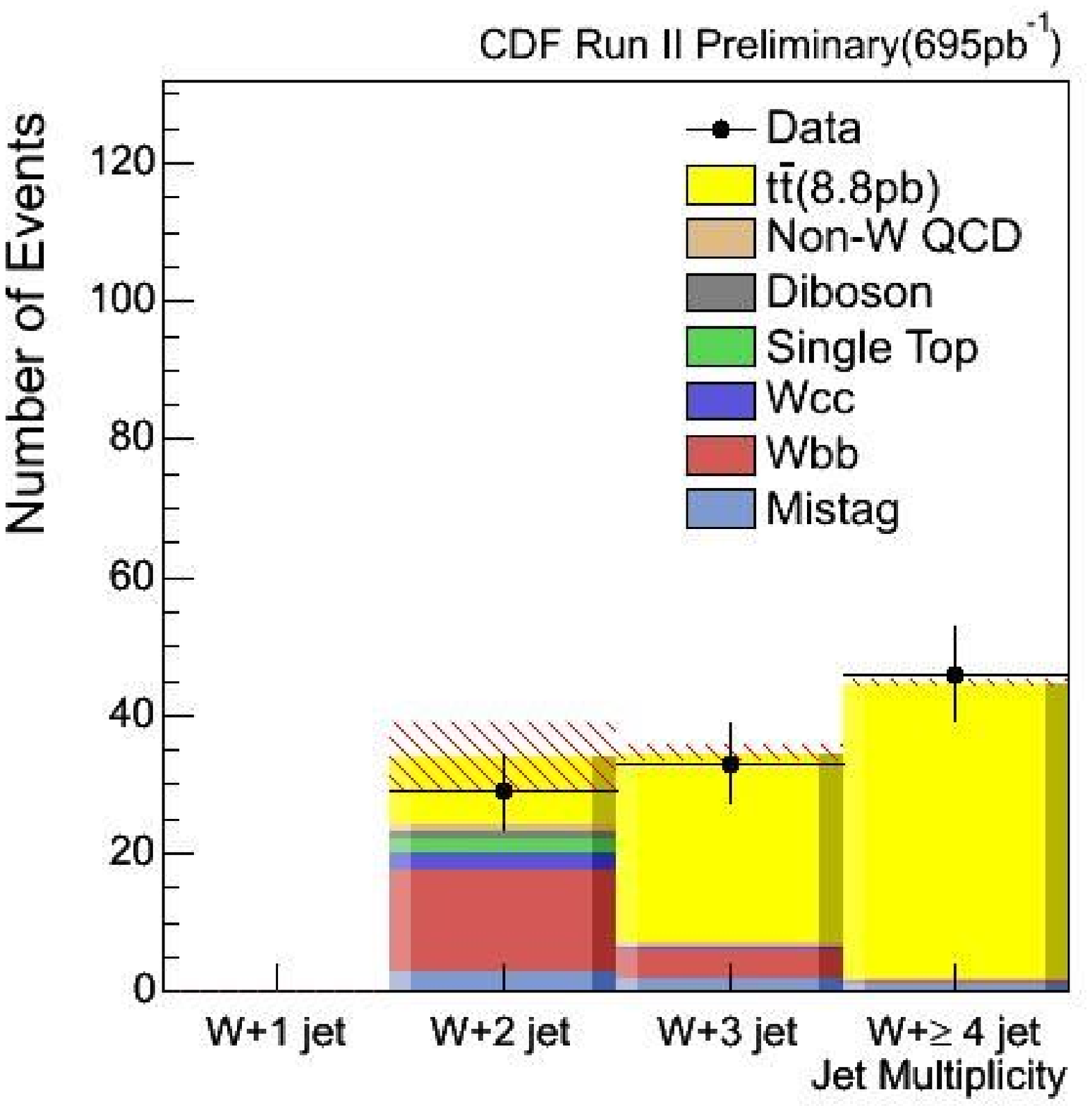}
}
\caption{\ttbar\ and background contributions to W+jets samples with
single (left) and double (right) b-tags.
\label{fig:cdf-ttxs-lj}}
\end{figure}

One of the best measurements of \ttbar\ cross section has been obtained
by CDF using the lepton+jets channel and 695 \invpb\ of data
\cite{cdf-ttxs-lj}.
While the traditional analyses in this channel have used selections
based on topological variables to enhance \ttbar\ signal,
this measurement employs b-tag information to reduce backgrounds.
Events are required to have one isolated electron or $\mu$ with
\pt$>20$ \gevc\ and \MET $>20$ GeV, 
at least three jets with \pt$>15$ \gevc\ within $|\eta|<2$ and
the total scalar sum of transverse energies of all objects in the event 
$> 200$ GeV (including jets with \pt$>8$ \gevc\ and $|\eta|<2.4$). 
The last requirement is dropped for the double-tagged sample.
As illustrated in Fig.~\ref{fig:cdf-ttxs-lj}(left), the events in
the ``W+3 jet'' and ``W+$\ge$4jet'' bins are relatively background
free and dominated by \ttbar\ contribution when one b-tag is required.
The resulting cross section is
$\sigma_{\rm t\bar{t}} = 8.2\pm 0.6{\rm (stat.)}\pm 1.0 {\rm (syst.)}$ pb.
The uncertainty is dominated by systematics, and its largest component
comes from b-tagging. 
When two b-tags are required, the sample statistics is reduced but \ttbar\
purity improves even further.
It is noteworthy that the cross section measurement using the
double-tagged sample alone has achieved a $5\sigma$ significance:
$\sigma_{\rm t\bar{t}} = 8.8^{+1.2}_{-1.1}{\rm (stat.)}^{+2.0}_{-1.3} {\rm (syst.)}$ pb.

\begin{figure}[!Hhtb]
\centerline{
\epsfxsize=2.2in
\epsffile{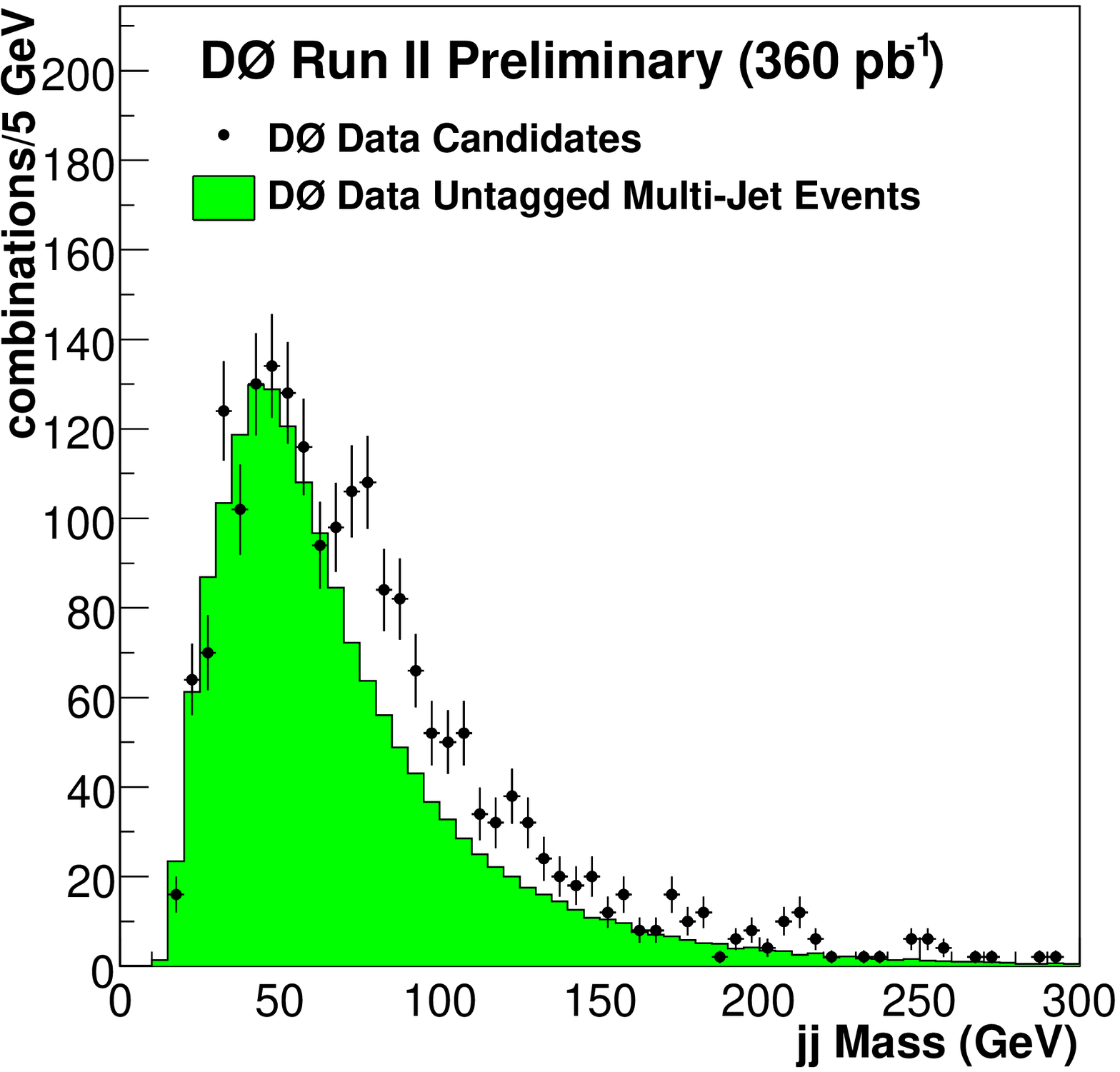}
\epsfxsize=2.2in
\epsffile{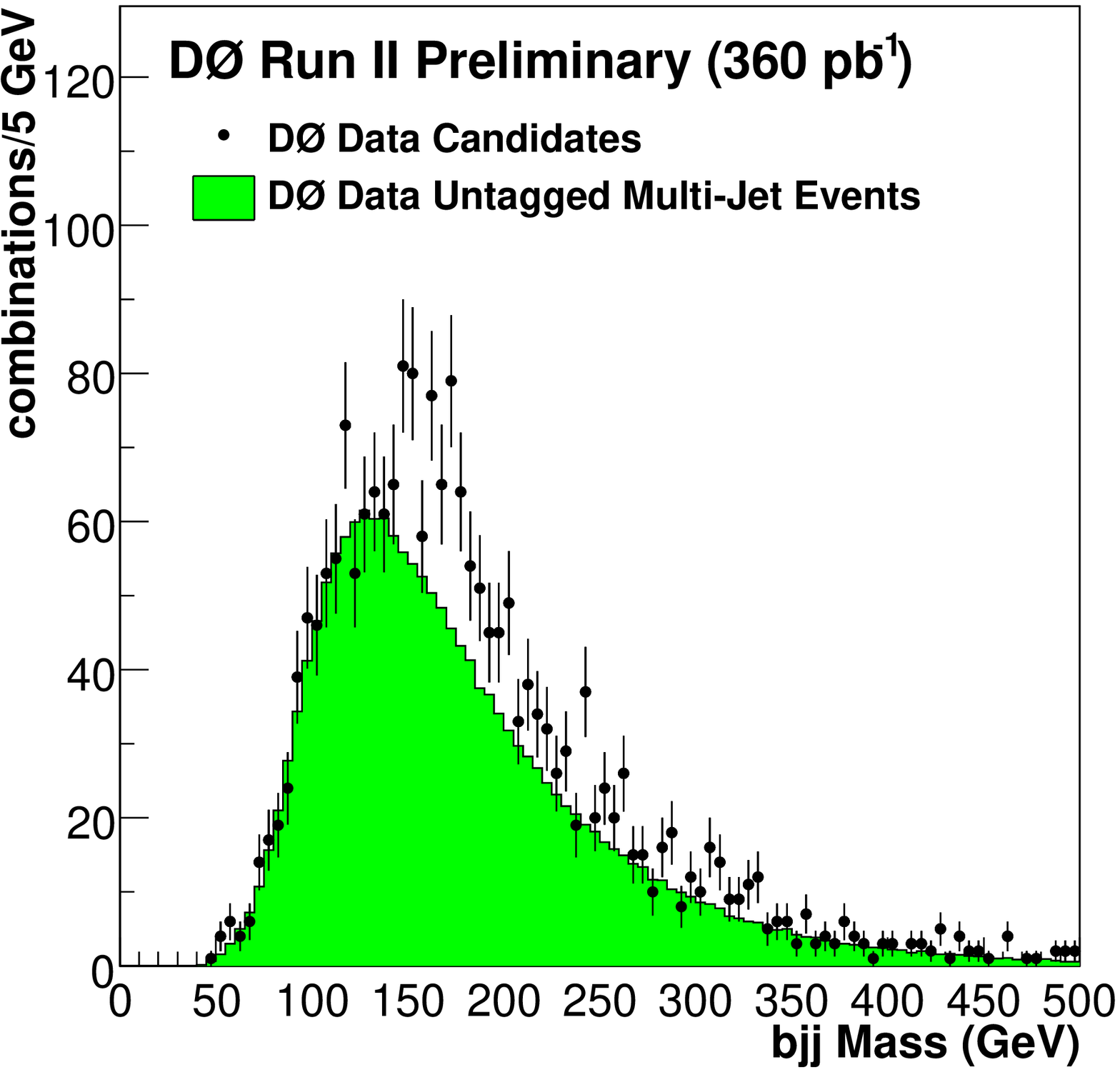}
}
\caption{The dijet mass for non b-tagged jets (jj) (left) and
the three-jet mass for one b-tagged and two non b-tagged jets (bjj) (right)
with non-\ttbar\ background overlayed. 
\label{fig:d0-ttxs-alljets}}
\end{figure}

CDF and D0 are developing a variety of techniques
to examine \ttbar\ decays into the all-hadronic final state.
A novel analysis from D0 \cite{d0-alljets}, based on 360 \invpb,
selects six-jet events with at least 2 jets 
having \pt$ > 45$ \gevc\ and tagged as b-jets with a secondary-vertex tagging
algorithm. The remaining jets are required not to be b-tagged, two of them
to have \pt$> 20$ \gevc\ and the rest \pt$ >15$ \gevc. All jets are
required to be within $|y|<2.4$. As no events have been rejected based on
the presence of high-\pt\ leptons or \MET, this sample includes contributions
from the all-hadronic channel, the $\tau$ channel with hadronic $\tau$
decays, and the other \ttbar\ decay channels when additional jets
are produced. The double b-tag requirement is essential for suppressing
the QCD backgrounds. The inclusive dijet mass distribution for non b-tagged
jets (jj), and the three-jet mass distribution for one b-tagged and two 
non-tagged jets (bjj) exhibit visible excess of events above
a smooth background. This enhancement is interpreted as due to W and top production 
(Fig.~\ref{fig:d0-ttxs-alljets}).
A method has been developed to derive the non-\ttbar\ background directly
from the data. After background subtraction the jj and bjj mass distributions 
agree well with expectations for  W and top decays into jets
based on {\sc pythia} and  simulation of detector effects.
The resulting \ttbar\ cross section of $12.1\pm 4.9\pm 4.6$ pb (for $m_{\rm t}=175$ GeV) 
is consistent with SM predictions. 
Its accuracy
is expected to be significantly improved when larger data samples are
analyzed and the technique is furter developed.
The direct observation of resonant W and top mass peaks in the hadronic
mode is reassuring in anticipation of the LHC data.

\begin{figure}[!Hhtb]
\centerline{
\epsfxsize=2.6in
\epsffile{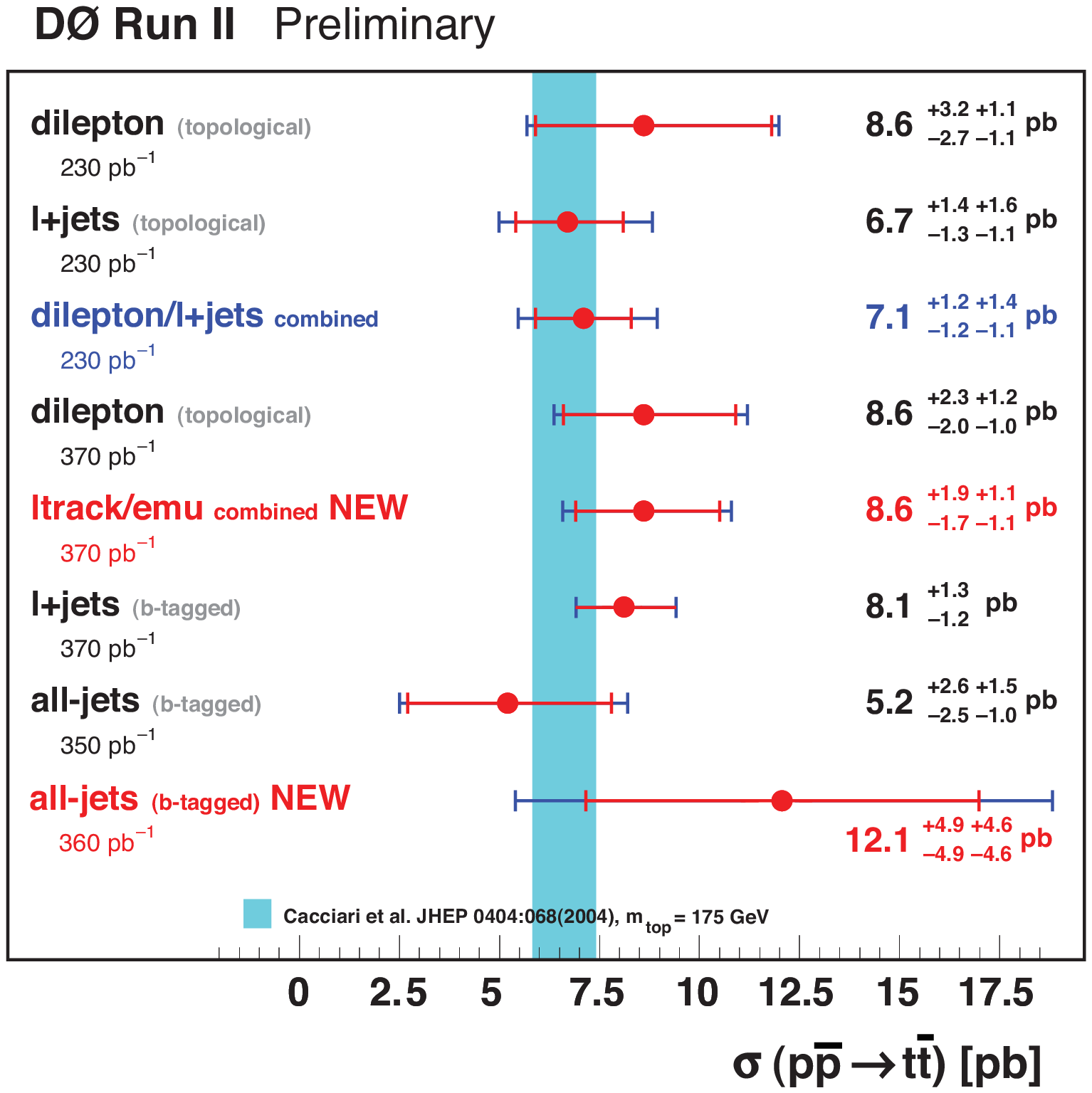}
\epsfxsize=1.8in
\epsffile{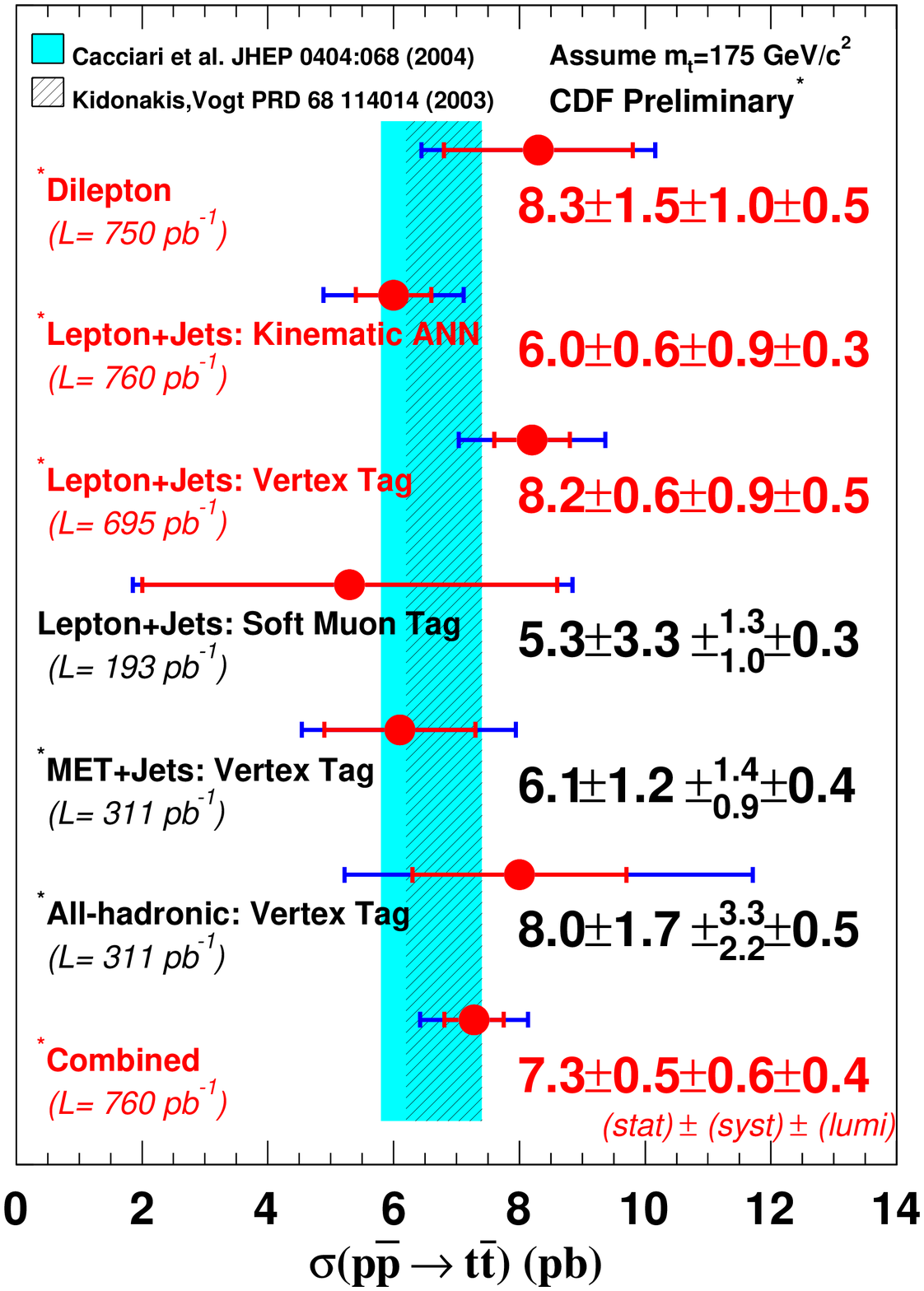}
}
\caption{Summary of recent \ttbar\ cross section results from D0 and CDF.
\label{fig:ttxs-summary}}
\end{figure}

Figure~\ref{fig:ttxs-summary} shows a summary of recent \ttbar\
cross section measurements by D0 and CDF. 
The accuracy of the combined result 
is approaching 10\%. With further increase 
of the data sets, it is becoming possible to test compatibility
of the cross sections obtained from different channels.

The top quark mass is a fundamental parameter of the SM and
should be measured to the highest possible accuracy.
With large data samples now available the measurements
are no longer statistically limited. It is therefore
important to understand systematic uncertainties in detail
and to minimize their impact on the determination of $m_{\rm t}$.
Since the dominant source of systematic uncertainty has been
the jet energy scale (JES), recent analyses employ the {\it in situ}
jet calibration by imposing the well known mass of the W 
in the reconstruction of the W$\rightarrow$jj decays in the \ttbar\
samples. This allows to further
constrain the overall JES in a
simultaneous fit to $m_{\rm t}$ and $m_{\rm W}$.

\begin{figure}[!Hhtb]
\centerline{
\epsfxsize=2.2in
\epsffile{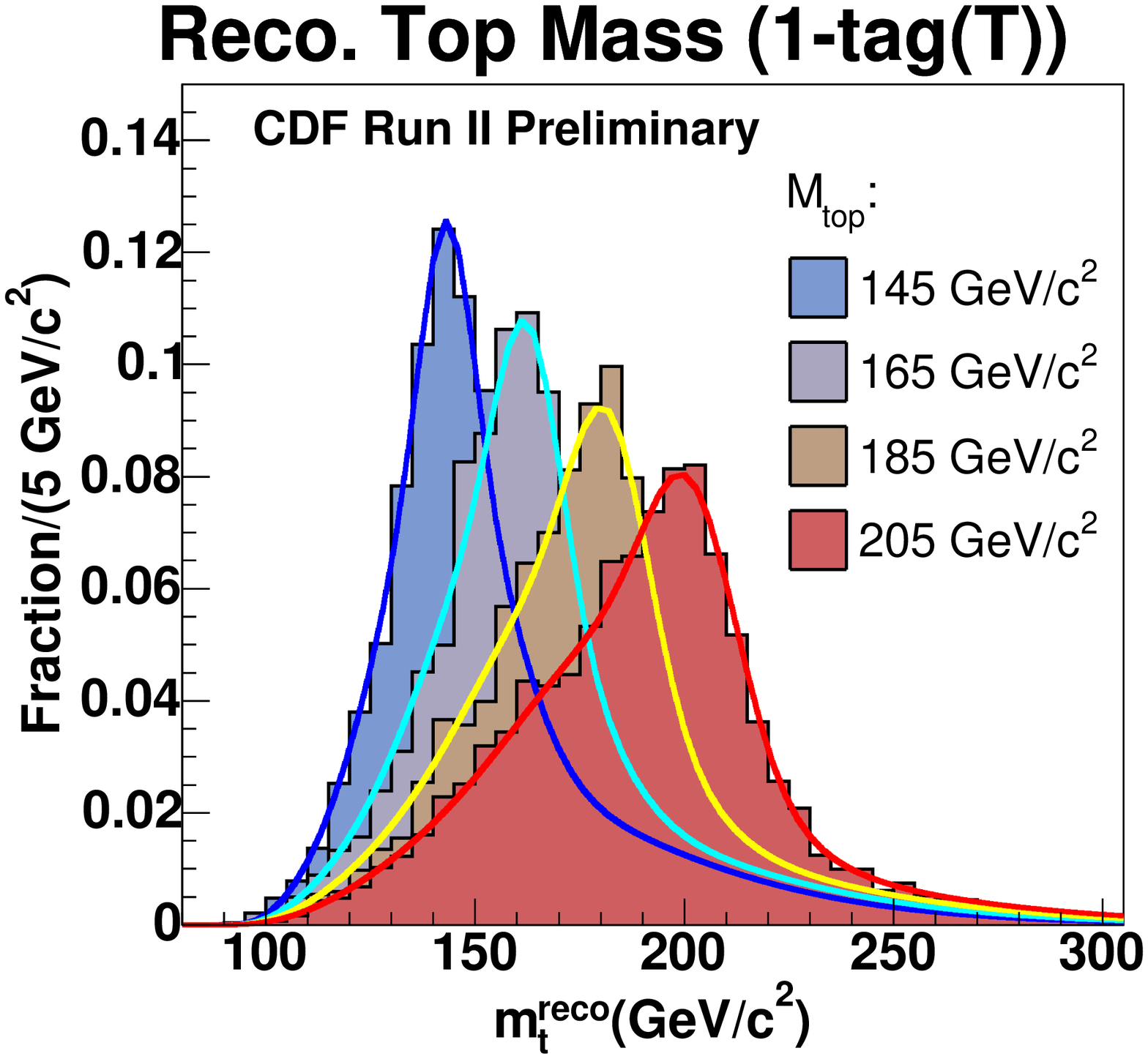}
\epsfxsize=2.8in
\epsffile{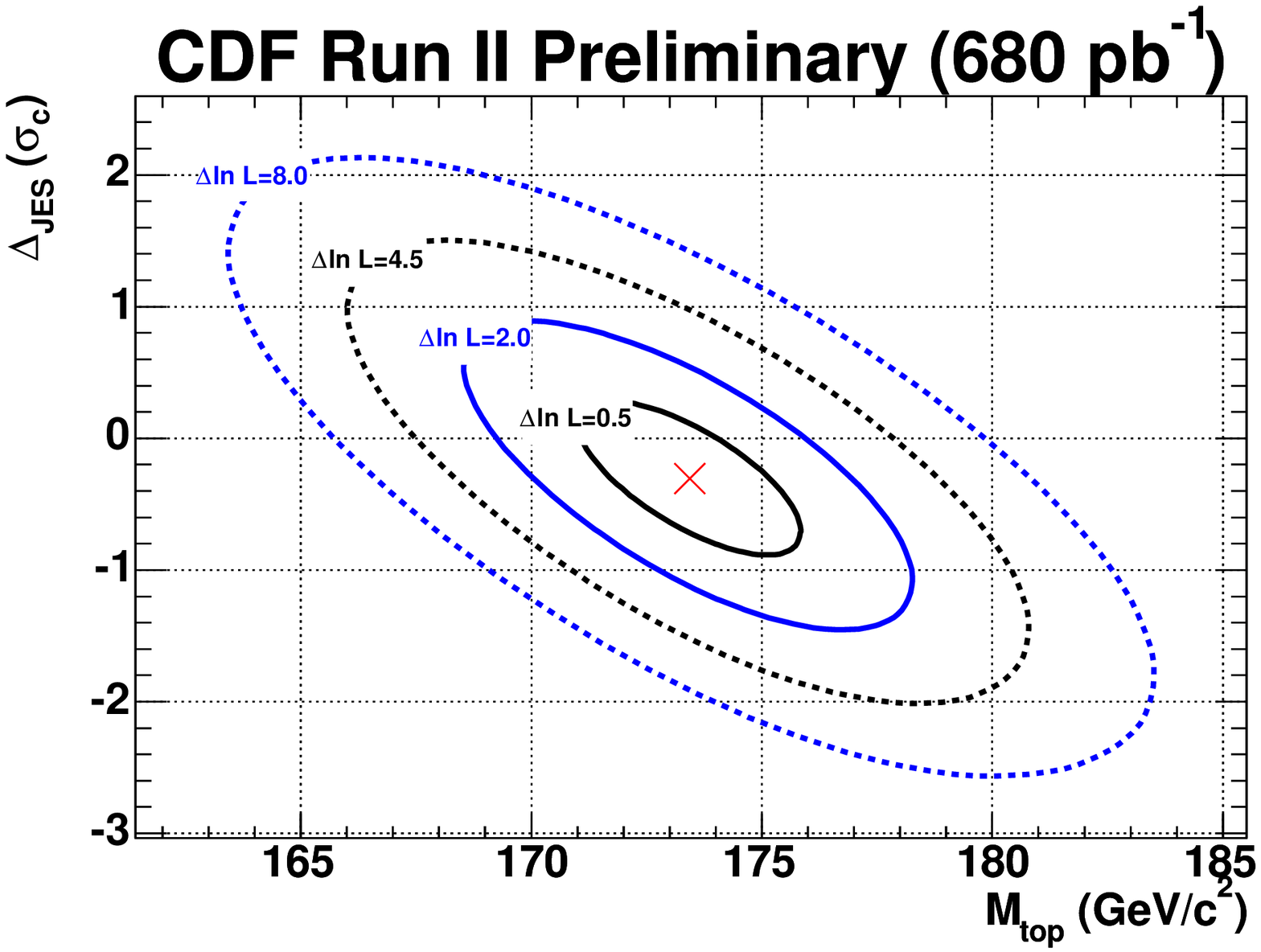}
}
\caption{
Left: $m_{\rm t}$ templates from Monte Carlo.
Right: Result of the template fit to lepton+jets data
vs $m_{\rm t}$ and $\Delta_{\rm JES}$.
\label{fig:cdf-top-mass-template}}
\end{figure}

CDF and D0 applied several sophisticated techniques in measurements of $m_{\rm t}$.
The major methods are ``template'' and ``matrix element'' approaches.
CDF performed the template analysis using lepton+jets channel and
680 \invpb\ \cite{cdf-top-mass-template}. The event sample
has been selected using
requirements similar to those described above for their
cross section measurement (with jet \pt\ cuts depending on the
b-tagging category of each event). A kinematic fit is used to 
decide the best value of $m_{\rm t}$ for each event after
considering all parton-to-jet assignments and constraining
the fitted  W mass to the book value.
The resulting  $m_{\rm t}$ distribution is then compared to
Monte Carlo $m_{\rm t}$ templates simulated for various
top masses as illustrated in Fig.~\ref{fig:cdf-top-mass-template}(left). 
The final reconstructed top mass is
determined from a simultaneous fit of the templates to the
observed distribution, as function of $m_{\rm t}$ and a shift
in the jet energy scale, $\Delta_{\rm JES}$, 
Fig.~\ref{fig:cdf-top-mass-template}(right).
The fit yields a top-quark mass of 
$m_{\rm t} = 173.4 \pm 2.8$ GeV.
The {\it in situ} calibration is consistent with the standard
calibration but reduces the JES-related uncertainty
by $\approx 40$\%. 

\begin{figure}[!Hhtb]
\centerline{
\epsfxsize=2.5in
\epsffile{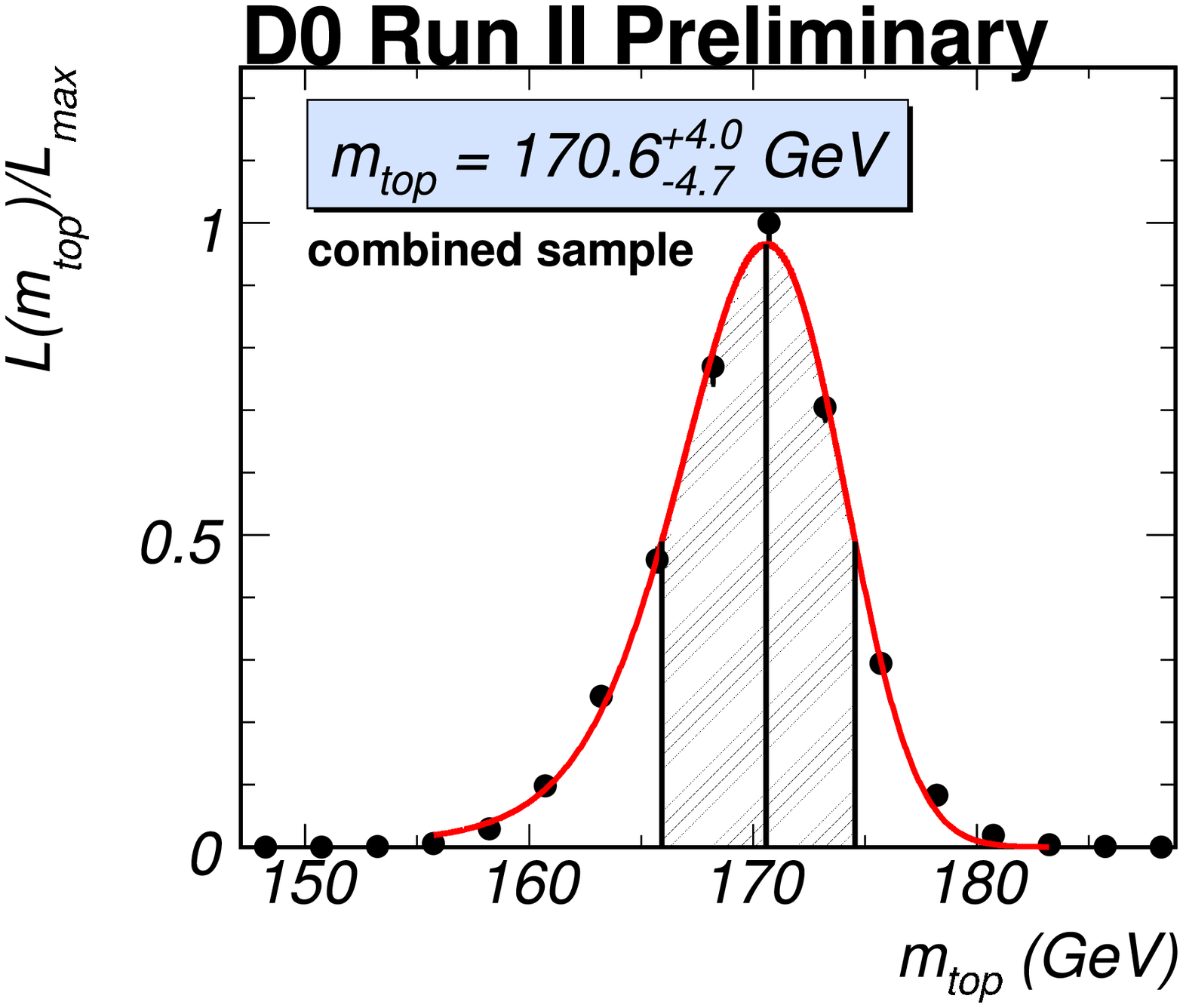}
\epsfxsize=2.5in
\epsffile{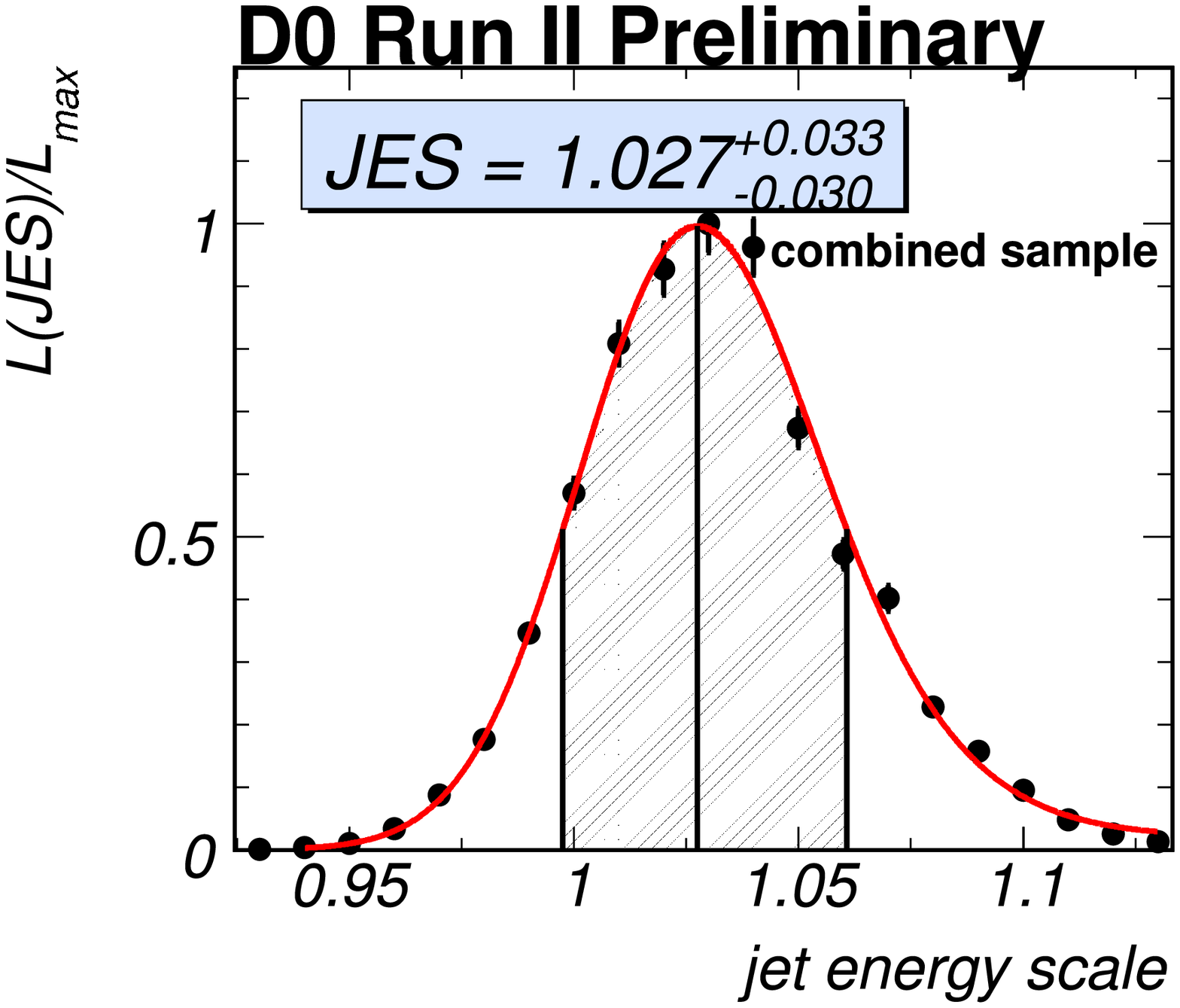}
}
\caption{Likelihood distributions vs $m_{\rm t}$ (left) and 
$\Delta_{\rm JES}$ (right), and the 68\% confidence-level (CL) intervals, 
using the ME method and b-tagging information.
\label{fig:d0-top-mass-ME}}
\end{figure}

D0 developed the matrix element (ME) method in Run 1
and applied it to 370 \invpb\ of Run 2 data
\cite{d0-top-mass-ME}.
In this method the probabilities for an event to be
\ttbar\ signal or the dominant W+jets background are 
calculated using the corresponding LO matrix elements. 
The probabilities of all events are combined into a
final likelihood, which is then maximized as a function of
$m_{\rm t}$ and an overall JES factor (in the Run 2
implementation). The likelihood distributions for both
parameters are shown in Fig.~\ref{fig:d0-top-mass-ME}.
The result using the b-tagging information is
$m_{\rm t} = 170.6^{+4.0}_{-4.7} {\rm (stat.+JES)} \pm 1.4 {\rm (syst.)}$ GeV.

\begin{figure}[!Hhtb]
\centerline{
\epsfxsize=1.6in
\epsffile{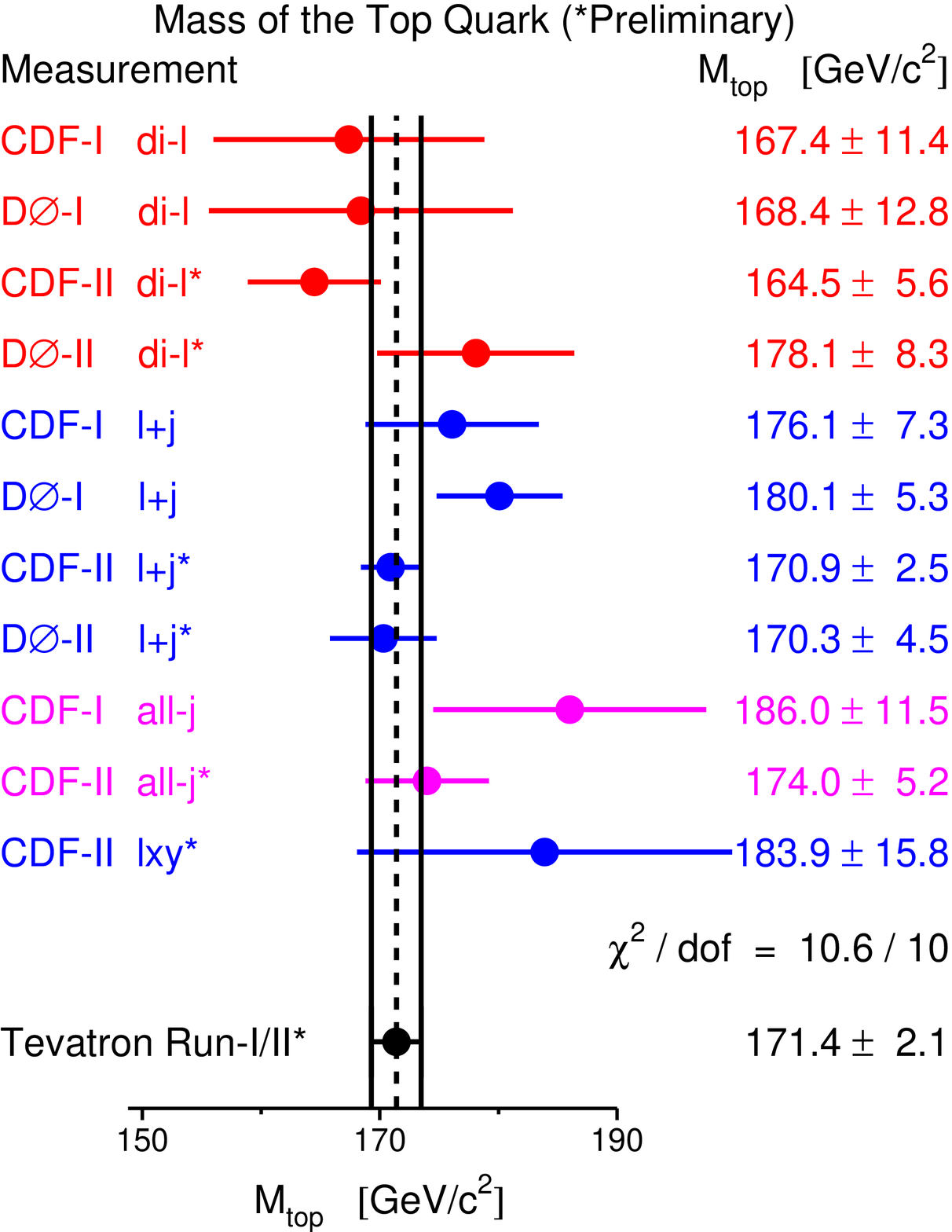}
\hfil
\epsfxsize=2.3in
\epsffile{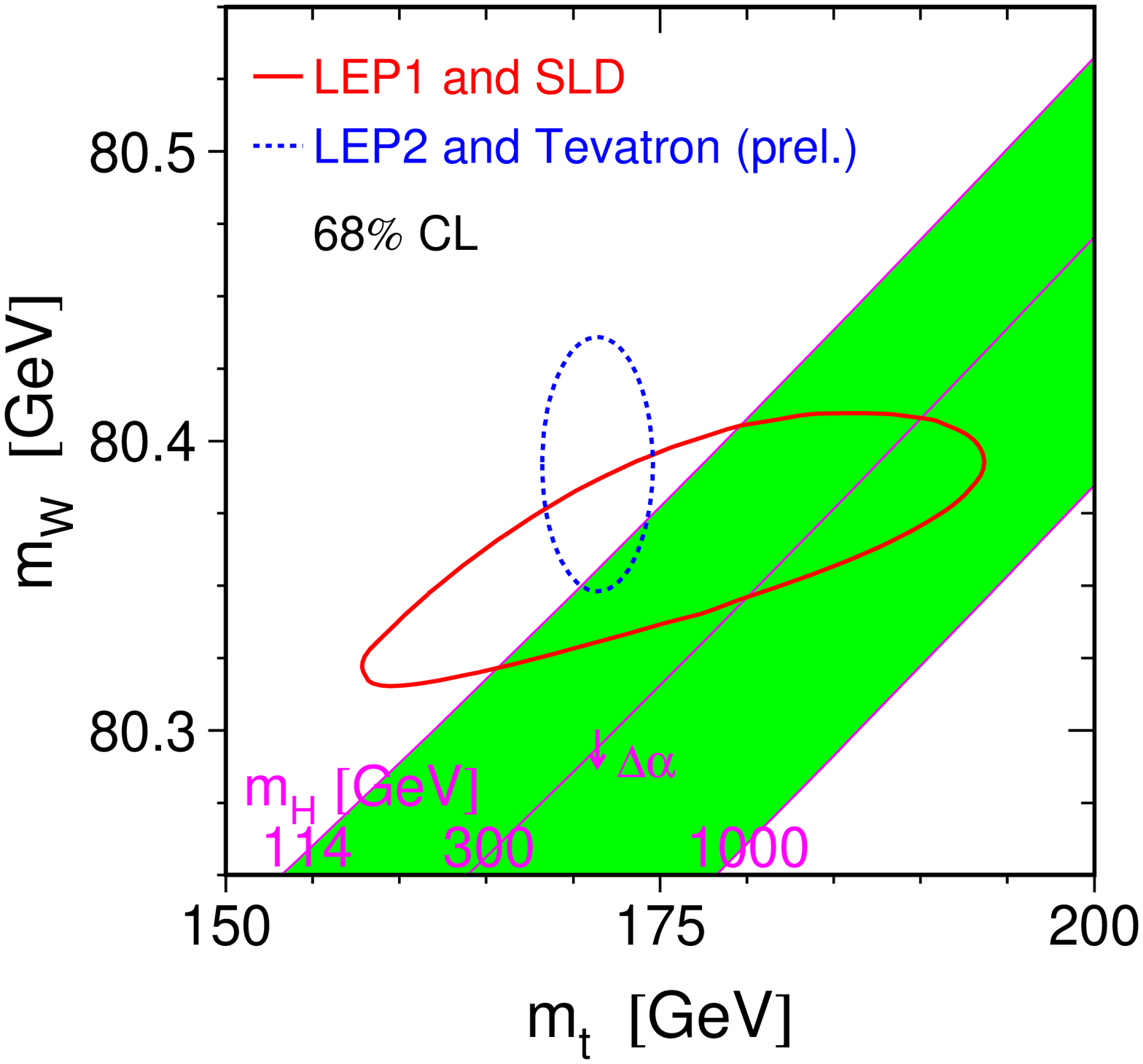}
}
\caption{
Left: Combination of best independent measurements of $m_{\rm t}$.
Right: Constraints on $m_{\rm H}$ from global electroweak SM fits
in the $m_{\rm t}$ and $m_{\rm W}$ plane.
\label{fig:top-mass-summary}}
\end{figure}

Figure~\ref{fig:top-mass-summary}(left) summarizes the best independent 
top-quark mass measurements from CDF and D0 \cite{top-mass-summary}. 
The combination of published Run 1 measurements with the recent 
preliminary Run 2 results using up to 1 \invfb\ of data yields a
preliminary world average mass of the top quark $m_t = 171.4 \pm 2.1$ GeV.
The top-quark mass is now known with a precision of 1.2\%.
The precise measurements of the top and W masses can be used to constrain
the value of $m_{\rm H}$, as illustrated in the right panel.
They suggest a low value of the mass of the Higgs boson 
setting the stage for an exciting race between Tevatron and 
LHC experiments towards its discovery.

\begin{figure}[!Hhtb]
\centerline{
\epsfxsize=2.5in
\epsffile{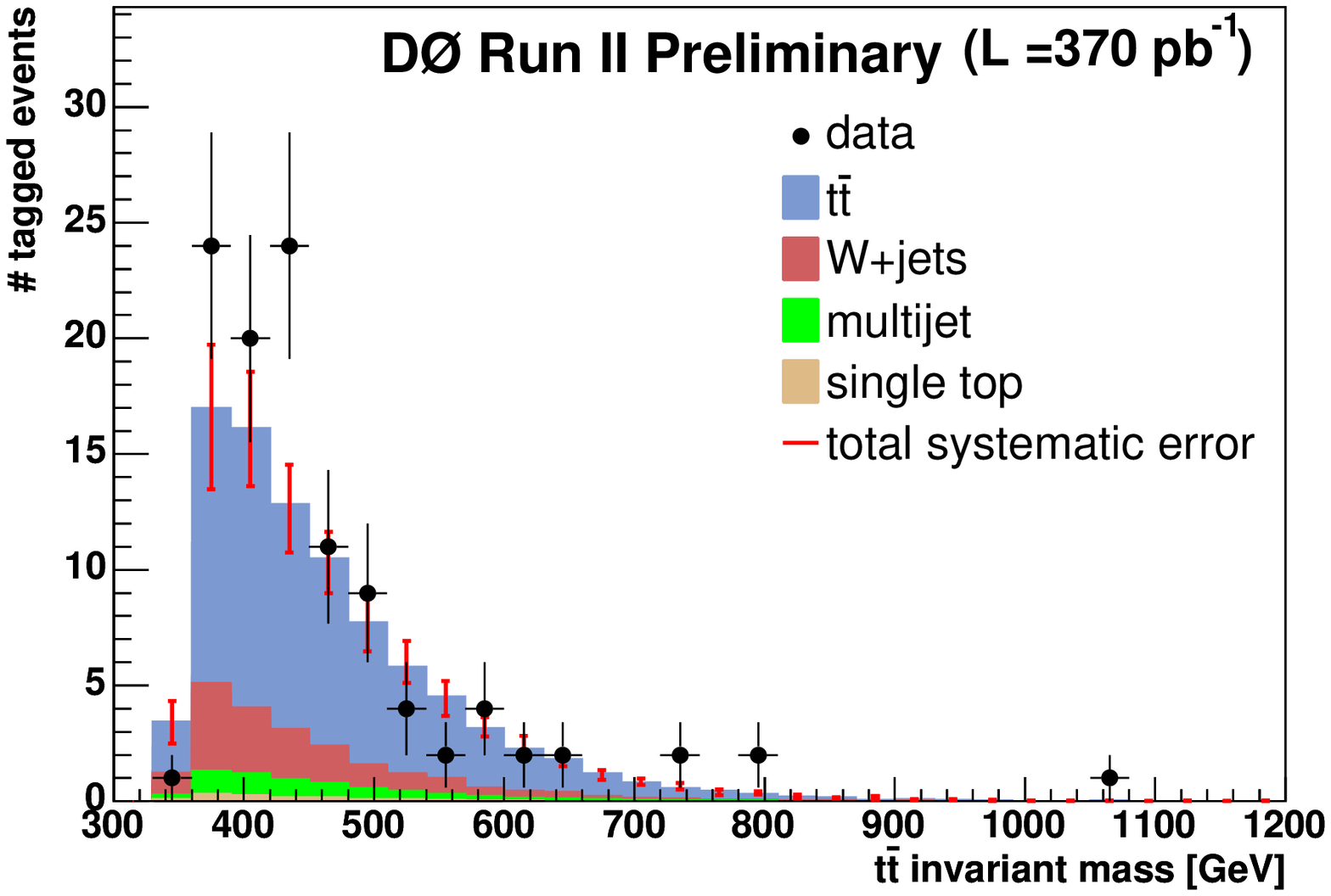}
\epsfxsize=2.5in
\epsffile{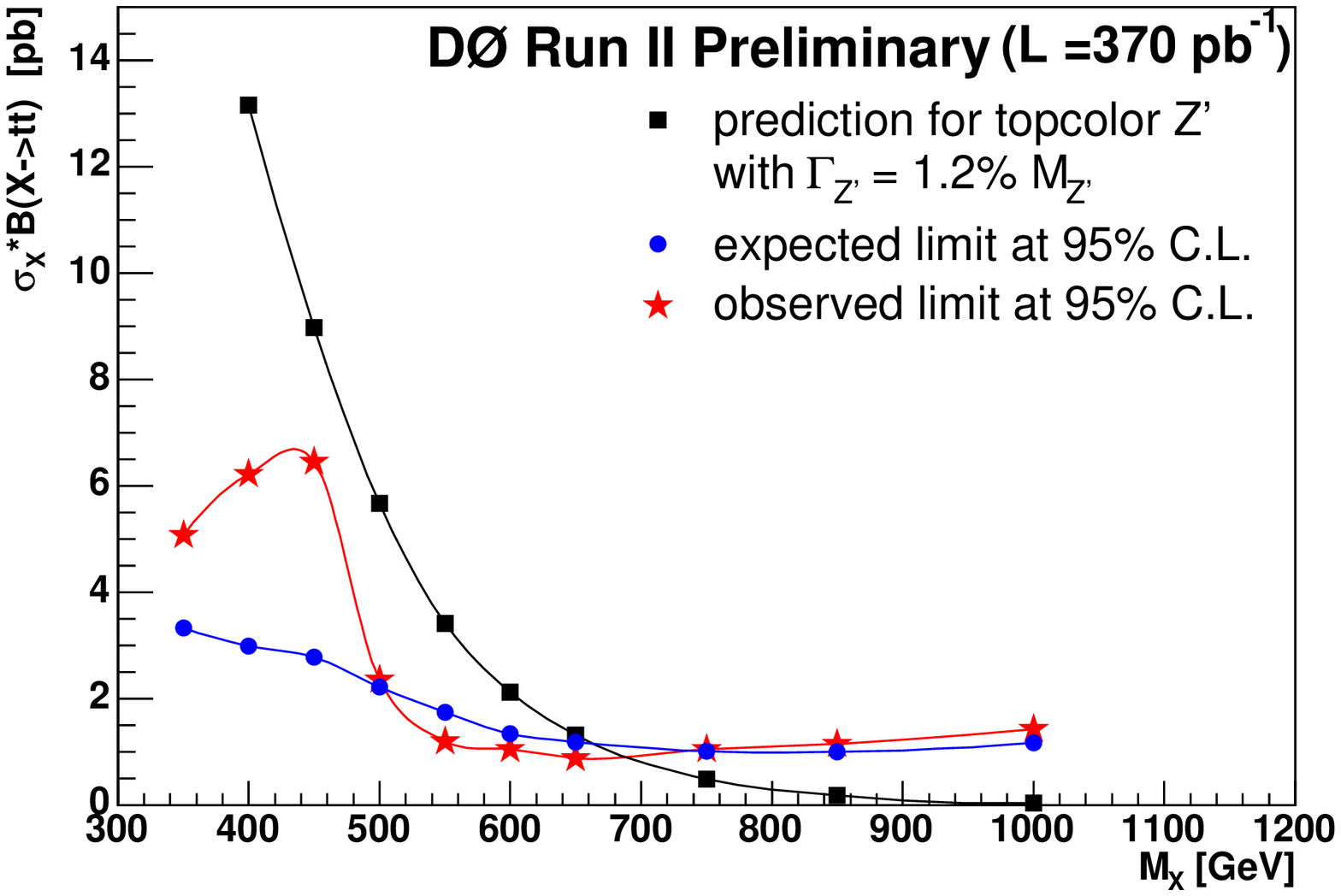}
}
\caption{
Left: \ttbar\ mass distribution in lepton+jets channel.
Right: 95\% CL upper limits on $\sigma_{\rm X}\times B({\rm X\rightarrow t\bar{t}})$
compared to a prediction for a topcolor Z'.
\label{fig:d0-tt-resonance}}
\end{figure}

D0 and CDF have searched for a narrow-width heavy resonance X decaying
into top-quark pairs \cite{tt-res}. 
Such resonant \ttbar\ production is expected 
eg. in various ``topcolor'' models.
The \ttbar\ invariant mass spectrum from D0
is shown in Fig.~\ref{fig:d0-tt-resonance}(left). This analysis is based on
lepton+jets channel using a lifetime tag to identify b-quarks in
370 \invpb of data. 
No evidence for a \ttbar\ resonance X was found
by either collaboration and upper limits on 
$\sigma_{\rm X}\times B({\rm X\rightarrow t\bar{t}})$ have been derived
as a function of $m_{\rm X}$ (Fig.~\ref{fig:d0-tt-resonance}(right)
for D0 results). For a topcolor Z' model \cite{topcolor},
the existence of a leptophobic Z' boson with mass $m_{\rm Z'} < 680$
(725) GeV has been excluded by D0 (CDF) at 95\% CL, for 
$\Gamma_{\rm Z'} = 0.012 m_{\rm Z'}$.

\begin{figure}[!Hhtb]
\centerline{
\epsfxsize=2.in
\epsffile{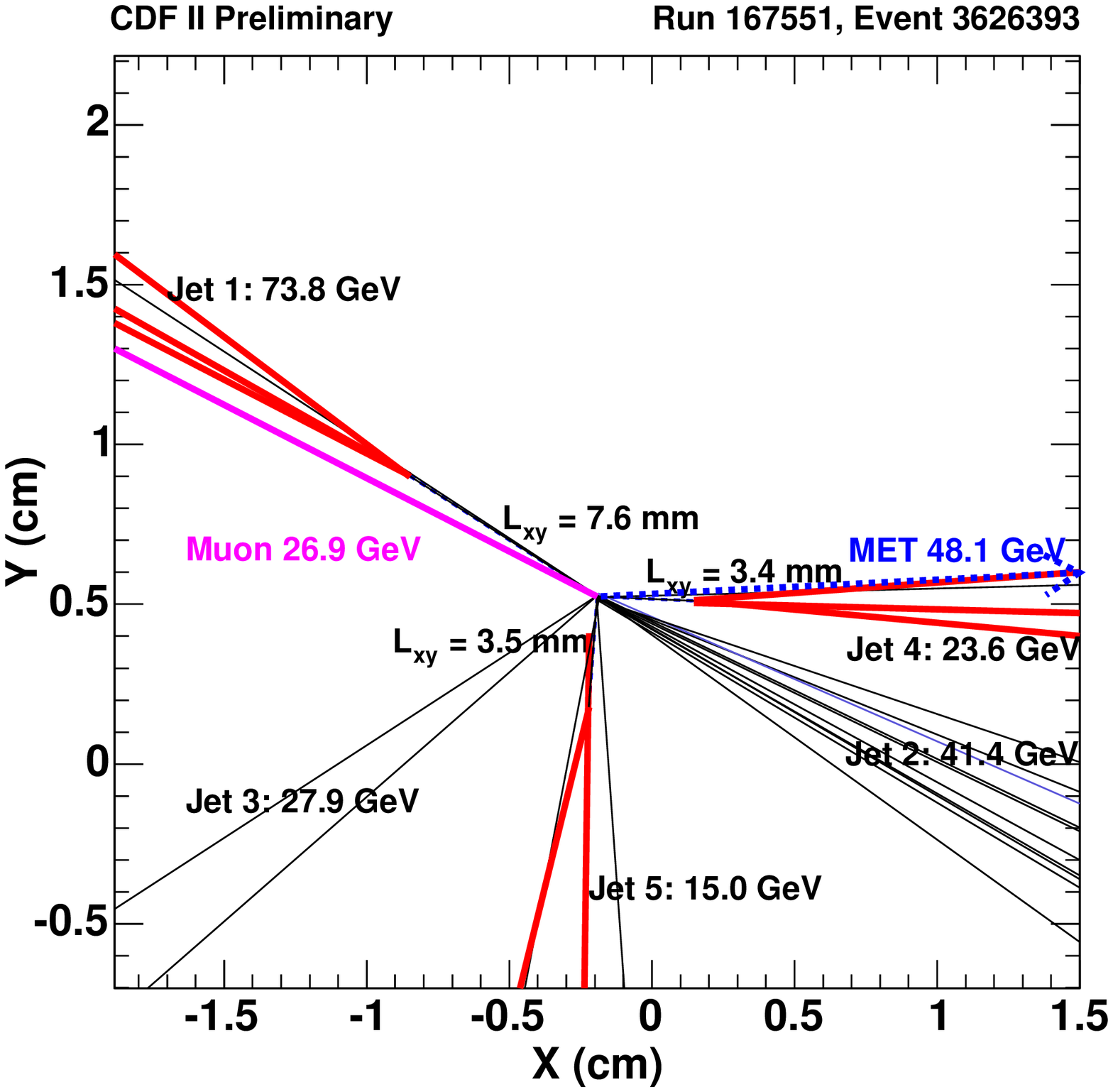}
\epsfxsize=2.7in
\epsffile{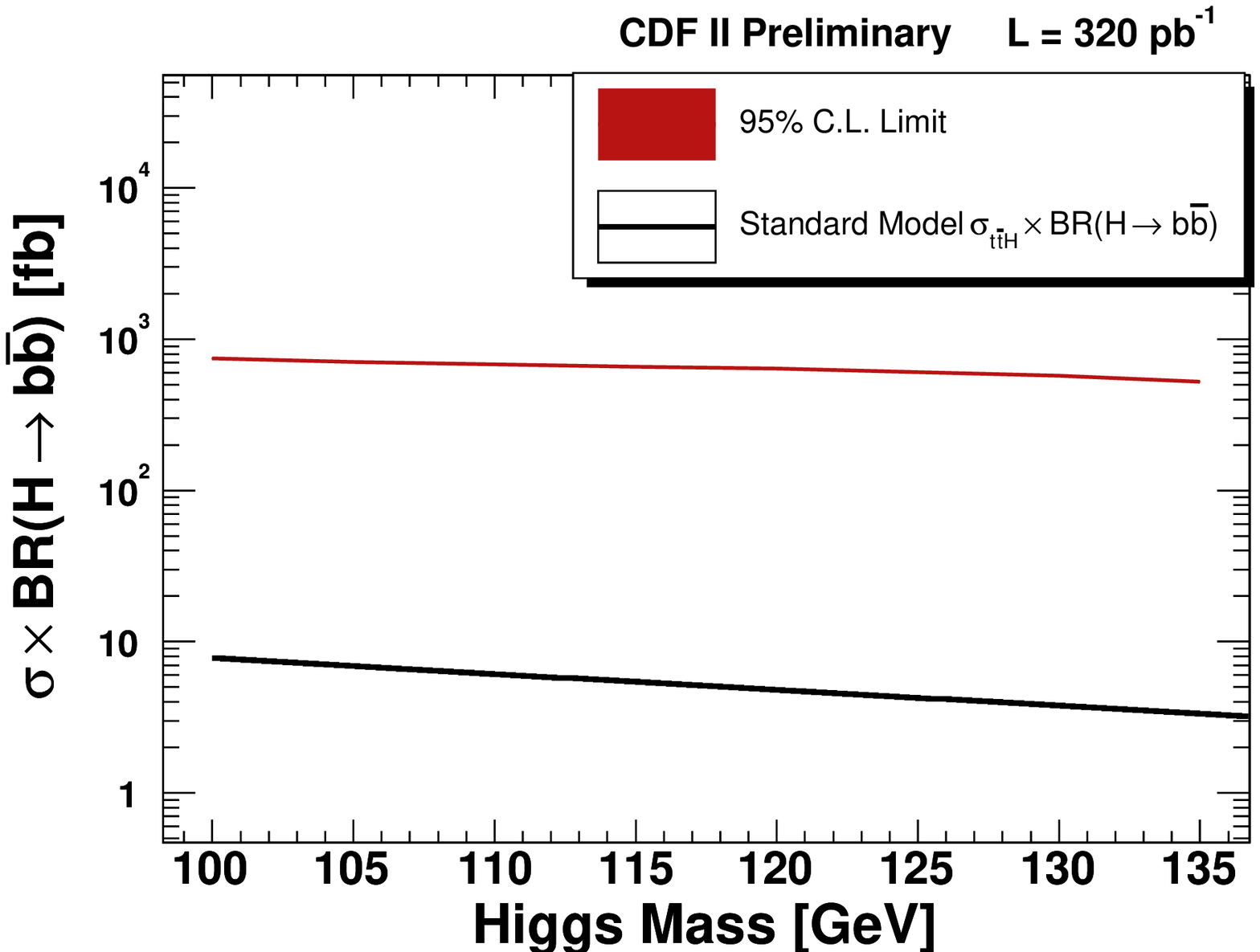}
}
\caption{
Left:  \ttbar H candidate event from CDF.
Right: 95\% CL upper limit and SM prediction for  
$\sigma_{\rm ttH}\times B({\rm H\rightarrow b\bar{b}})$ vs $m_{\rm H}$.
\label{fig:cdf-ttH}}
\end{figure}

Using 320 \invpb, CDF searched \cite{cdf-ttH}
for the W$^+$W$^-$\bbbar\bbbar\ signature of the associated \ttbar H
production. This process is expected to help the discovery
of a light Higgs and provide a determination of the t-H coupling at the LHC.
The CDF analysis required
a \pt$>20$ \gevc\ e or $\mu$ candidate, five or more jets
with $E_T >15$ GeV and $|\eta|< 2$, three or more b-tagged jets, 
and \MET $>10$ GeV. One candidate event was found (Fig.~\ref{fig:cdf-ttH}(left)), 
consistent with
the total expected background of $0.89\pm 0.12$ events. The major
contributions to background were from mistagging a light-quark jet as
a b-jet, QCD multijet events where a jet fakes a lepton, and irreducible
backgrounds from SM sources (including \ttbar\bbbar, \ttbar\ccbar\ etc.).
CDF obtained the first experimental limit 
on $\sigma_{\rm ttH}\times B({\rm H\rightarrow b\bar{b}})$
of 660 fb, weakly depending on $m_{\rm H}$ (Fig.~\ref{fig:cdf-ttH}(right)).
The expected \ttbar H signal is $0.024\pm 0.005$ events
for $m_{\rm H} = 115$ GeV.

\section{Conclusions}

Tevatron measurements advance the understanding of
``soft'' and ``hard'' aspects of QCD including higher-order
processes and multi-jet radiation; facilitate development
and tuning of perturbative and Monte Carlo tools; improve understanding
of PDFs, jet algorithms and calibrations. Building upon 
this progress and using large data samples that have become available
in Run 2, top-quark studies have entered a precision era, 
providing determination of \ttbar\ cross section approaching 10\% precision 
and of the top mass nearing 1\%.
Advanced analysis methods have been developed and
tried in the hadron-collider environment.
The experience from the Tevatron is an extremely valuable resource
and it can greatly benefit
the ``rediscovery'' of the Standard Model and searches for new
physics at the LHC.

\end{document}